\documentclass[10pt]{article}
\usepackage{amsmath, amsthm, amssymb}
\usepackage{float,epsfig}
\usepackage{tkz-graph}
\usepackage{color}
\usepackage{braket}
\usepackage{algorithm}
\usepackage[utf8]{inputenc}
\usepackage{tikz}
\usepackage{palatino}
\usetikzlibrary{quantikz}
\usepackage{cancel}
\usepackage{graphicx}
\usepackage{rotating}
\usepackage{lscape}
\usepackage{caption}
\usepackage{subcaption}
\usepackage{tikz}
\usepackage{booktabs}
\usepackage{algpseudocode}
\usepackage{algorithm}
\usepackage{hyperref}
\usepackage{cleveref}
\usepackage{comment}
\usetikzlibrary{arrows}
\usetikzlibrary{patterns}

\begin{document}

\newtheorem{theorem}{\bf Theorem}[section]
\newtheorem{proposition}[theorem]{\bf Proposition}
\newtheorem{definition}[theorem]{\bf Definition}
\newtheorem{corollary}[theorem]{\bf Corollary}
\newtheorem{example}[theorem]{\bf Example}
\newtheorem{exam}[theorem]{\bf Example}
\newtheorem{remark}[theorem]{\bf Remark}
\newtheorem{lemma}[theorem]{\bf Lemma}
\newcommand{\nrm}[1]{|\!|\!| {#1} |\!|\!|}

\newcommand{\calL}{{\mathcal L}}
\newcommand{\calX}{{\mathcal X}}
\newcommand{\calA}{{\mathcal A}}
\newcommand{\calB}{{\mathcal B}}
\newcommand{\calC}{{\mathcal C}}
\newcommand{\calK}{{\mathcal K}}
\newcommand{\C}{{\mathbb C}}
\newcommand{\R}{{\mathbb R}}
\renewcommand{\SS}{{\mathbb S}}
\newcommand{\LL}{{\mathbb L}}
\newcommand{\st}{{\star}}
\def\kernel{\mathop{\rm kernel}\nolimits}
\def\sigan{\mathop{\rm span}\nolimits}

\newcommand{\klasse}{{\boldsymbol \Delta}}

\newcommand{\ba}{\begin{array}}
\newcommand{\ea}{\end{array}}
\newcommand{\von}{\vskip 1ex}
\newcommand{\vone}{\vskip 2ex}
\newcommand{\vtwo}{\vskip 4ex}
\newcommand{\dm}[1]{ {\displaystyle{#1} } }

\newcommand{\be}{\begin{equation}}
\newcommand{\ee}{\end{equation}}
\newcommand{\beano}{\begin{eqnarray*}}
\newcommand{\eeano}{\end{eqnarray*}}
\newcommand{\inp}[2]{\langle {#1} ,\,{#2} \rangle}
\def\bmatrix#1{\left[ \begin{matrix} #1 \end{matrix} \right]}
\def \noin{\noindent}
\newcommand{\evenindex}{\Pi_e}

%\newcommand {\proof} {\par{\it Proof}. \ignorespaces}
%\newcommand {\eproof}
%      {\sigace
%        {\ \vbox{\hrule\hbox{\vrule height1.3ex\hskip0.8ex\vrule}\hrule}}
%        \par}

%%%%%%%%%%%%%%%%%%%%%%%%%%%%%%%%%%%%%%%%%%%%%%%%%%%%%%%%%%%%%%%%%%%%%%%%%%

\def \R{{\mathbb R}}
\def \C{{\mathbb C}}
\def \K{{\mathbb K}}
\def \H{{\mathbb H}}

\def \T{{\mathbb T}}
\def \Pb{\mathrm{P}}
\def \N{{\mathbb N}}
\def \Ib{\mathrm{I}}
\def \Ls{{\Lambda}_{m-1}}
\def \Gb{\mathrm{G}}
\def \Hb{\mathrm{H}}
\def \Lam{{\Lambda}}

\def \Qb{\mathrm{Q}}
\def \Rb{\mathrm{R}}
\def \Mb{\mathrm{M}}
\def \norm{\nrm{\cdot}\equiv \nrm{\cdot}}

\def \P{{\mathbb{P}}_m(\C^{n\times n})}
\def \A{{{\mathbb P}_1(\C^{n\times n})}}
\def \H{{\mathbb H}}
\def \L{{\mathbb L}}
\def \G{{\F_{\tt{H}}}}
\def \S{\mathbb{S}}
\def \s{\mathbb{s}}
\def \sigmin{\sigma_{\min}}
\def \elam{\Lambda_{\epsilon}}
\def \slam{\Lambda^{\S}_{\epsilon}}
\def \Ib{\mathrm{I}}
\def \Tb{\mathrm{T}}
\def \d{{\delta}}

\def \QT{{\texttt{QT}}}

\def \Lb{\mathrm{L}}
\def \N{{\mathbb N}}
\def \Ls{{\Lambda}_{m-1}}
\def \Gb{\mathrm{G}}
\def \Hb{\mathrm{H}}
\def \Delta{\triangle}
\def \Rar{\Rightarrow}
\def \p{{\mathsf{p}(\lam; v)}}

\def \D{{\mathbb D}}

\def \tr{\mathrm{Tr}}
\def \cond{\mathrm{cond}}
\def \lam{\lambda}
\def \sig{\sigma}
\def \sign{\mathrm{sign}}

\def \ep{\epsilon}
\def \diag{\mathrm{diag}}
\def \rev{\mathrm{rev}}
\def \vec{\mathrm{vec}}

\def \ham{\mathsf{Ham}}
\def \herm{\mathsf{Herm}}
\def \sym{\mathsf{sym}}
\def \odd{\mathsf{sym}}
\def \en{\mathrm{even}}
\def \rank{\mathrm{rank}}
\def \pf{{\bf Proof: }}
\def \dist{\mathrm{dist}}
\def \rar{\rightarrow}

\def \rank{\mathrm{rank}}
\def \pf{{\bf Proof: }}
\def \dist{\mathrm{dist}}
\def \Re{\mathsf{Re}}
\def \Im{\mathsf{Im}}
\def \re{\mathsf{re}}
\def \im{\mathsf{im}}

\def \mf{\mathcal{F}}
\def \ot{\overline{\theta}}
\def \od{\overline{\delta}}

\def \sym{\mathsf{sym}}
\def \sksym{\mathsf{skew\mbox{-}sym}}
\def \odd{\mathrm{odd}}
\def \even{\mathrm{even}}
\def \herm{\mathsf{Herm}}
\def \skherm{\mathsf{skew\mbox{-}Herm}}
\def \str{\mathrm{ Struct}}
\def \eproof{$\blacksquare$}

\def \bS{{\bf S}}
\def \cA{{\cal A}}
\def \E{{\mathcal E}}
\def \X{{\mathcal X}}
\def \F{{\mathcal F}}
\def \cH{\mathcal{H}}
\def \cJ{\mathcal{J}}
\def \tr{\mathrm{Tr}}
\def \range{\mathrm{Range}}
\def \adj{\star}
%\newcommand {\proof} {\par{\it Proof}. \ignorespaces}
%\newcommand {\eproof}
    %  {\sigace
        %{\ \vbox{\hrule\hbox{\vrule height1.3ex\hskip0.8ex\vrule}\hrule}}
        %\par}

\def \pal{\mathrm{palindromic}}
\def \palpen{\mathrm{palindromic~~ pencil}}
\def \palpoly{\mathrm{palindromic~~ polynomial}}
\def \odd{\mathrm{odd}}
\def \even{\mathrm{even}}

\algrenewcommand\algorithmicrequire{\textbf{Input:}}
\algrenewcommand\algorithmicensure{\textbf{Output:}}

\newcommand{\tm}[1]{\textcolor{magenta}{ #1}}
\newcommand{\tre}[1]{\textcolor{red}{ #1}}
\newcommand{\tb}[1]{\textcolor{blue}{ #1}}

%%%%%%%%%%%%%%%%%%%%%%%%%%%%%%%%%%%%%%%%%%%%%%%%%%%%%%%%%%%%%%%%%%%%%%%%%%%%%%%%

\title{Local Hamiltonian decomposition and classical simulation of  parametrized quantum circuits}
\author{Bibhas Adhikari\thanks{Corresponding author, Fujitsu Research of America, Inc., Santa Clara, CA, USA, Email: badhikari@fujitsu.com or bibhas.adhikari@gmail.com } \,\, and \,\,  Aryan Jha\thanks{Undergraduate student, Department of Mathematics, IIT Kharagpur, India, Email: aryanjha026@gmail.com } }

\date{}

\maketitle
\thispagestyle{empty}

\noindent{\bf Abstract.} In this paper we develop a classical algorithm of complexity $O(K \, 2^n)$ to simulate parametrized quantum circuits (PQCs) of $n$ qubits, where $K$ is the total number of one-qubit and two-qubit control gates. The algorithm is developed by finding $2$-sparse unitary matrices of order $2^n$ explicitly corresponding to any single-qubit and two-qubit control gates in an $n$-qubit system. Finally, we determine analytical expression of Hamiltonians for any such gate and consequently a local Hamiltonian decomposition of any PQC is obtained. All results are validated with numerical simulations. \\

%We derive   Hamiltonians for  two-qubit and string of one-qubit  quantum gates which constitute parametrized quantum circuits for multi-qubit systems. This problem is equivalent to finding a Hermitian matrix $H$ of order $2^n$ for a given sparse parametric unitary matrix $U(\theta)$ which corresponds to a one-qubit or two-qubit gate in a circuit of an $n$-qubit system such that $U(\theta)=\exp(-iH(\theta)),$ which leads to a decomposition of any PQC.  Further, the parametric expression of the unitary matrices $U$ is employed to derive analytical formula of probability amplitudes of the output quantum state for any input state of an $n$-qubit quantum circuit under measurement with respect to the computational basis. The obtained results are corroborated with numerical examples.   \\


%\noindent{\bf AMS subject classification(2000):} 

\section{Introduction}

An open problem in Quantum Machine Learning (QML) is to determine an efficient ansatz which is employed for optimization of loss function for a quantum variational algorithm (VQA) on Noisy Intermediate-Scale Quantum (NISQ) computers. Several notions are introduced in the literature recently to characterize ansatze and to study its influence on the performance of the VQAs. Ansatze are also known as Parametrized Quantum Circuits (PQCs) which are also considered as machine learning models \cite{benedetti2019parameterized} and form an integral part of  Quantum Neural Networks (QNNs) \cite{abbas2021power}. One of the fundamental issues in the study of ansatze as parametrized quantum circuits (PQCs) is training the parameters that often is hindered by the adverse effect of barren plateau phenomena, which is concerned with vanishing of gradient of the loss function of the associated optimization function. The quantum landscape theory addresses the study of loss function landscape and its impact on the optimization process \cite{ragone2023unified}.

In general, analysis of a PQC is performed by considering it as a sequence of parametrized single-qubit gates and two-qubit controlled parametrized gates. A multi-layer periodic PQC is composed of multiple, say $L$ identical layers of PQCs each of which represents a unitary matrix $U(\boldsymbol{\theta}^l),$ $l=1,\hdots,L,$ where $\boldsymbol{\theta}^l=\{\theta_{lk}\}_{k=1}^K$ is a set of parameters which defines a single layer, and $U(\boldsymbol{\theta}^l)=\prod_{k=1}^K e^{-\iota \theta_{lk}H_k},$  $\iota=\sqrt{-1},$ $H_k$ is a trace-less Hermitian matrix. Properties of PQCs depend on the properties of the matrices $H_k, 1\leq k\leq K,$ also known as \textit{local Hamiltonians} corresponding to the unitary evolution associated with the PQCs. For instance, barren plateau induced by overparametrization phenomenon in QNNs with periodic structure is investigated through the dynamical Lie algebra defined by commutators of the operators $H_k,$ also called  generators for the unitary matrix $U(\boldsymbol{\theta}^l)$ \cite{larocca2023theory,larocca2022diagnosing,goh2023lie}. The limitation of noisy VQAs is demonstrated by using the properties of $H_k$s in \cite{wang2021noise}.

\begin{figure}[!ht]
 \centering
    {\includegraphics[width=0.60\textwidth]{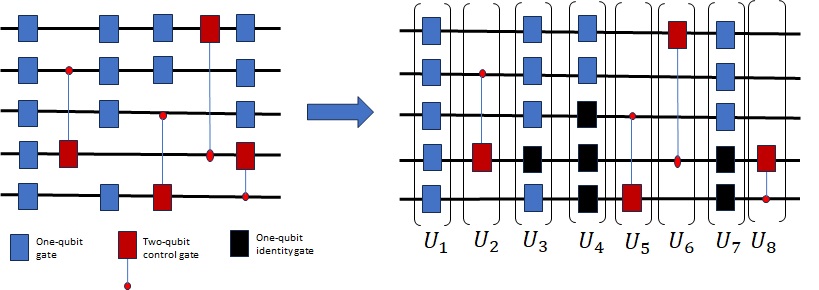} }%
 \centering \caption{A single layer parametric quantum circuit and its equivalent quantum circuit with separated two-qubit control gates and string of single-qubit control gates}
 \label{fig:PQC}
\end{figure}

It may further be noted that quantum circuits for $n$-qubit systems, as mentioned above, are usually analyzed by considering each layer of it as $$U(\boldsymbol{\theta})=\prod_{k=1}^K e^{-\iota \theta_kH_k},$$ where $\boldsymbol{\theta}=\{\theta_k\},$ the set of all the parameters, $\theta_k\in\R$, and $H_k$ is a Hermitian matrix of order $2^n$. Obviously, the QAOA circuit has this form, where two types of Hamiltonians are used, known as cost Hamiltonian and mixer Hamiltonian \cite{farhi2014quantum}. The main contribution of this paper is to show that the representation of any general one-layer quantum circuit is given by \begin{equation}\label{eqn:qcdecom}
    U(\boldsymbol{\theta})=\prod_{k=1}^K e^{-\iota\sum_{p=1}^P \lambda{pk}(\boldsymbol{\theta}_k)H_{pk}(\boldsymbol{\theta}_k)},
\end{equation} where $\lam_{pk}$ is a real-valued function, $\boldsymbol{\theta}=\{\boldsymbol{\theta}_k\},$ $\boldsymbol{\theta}_k=\{\theta_{mk}\}_{m=1}^M$, $\theta_{mk}\in\R$ for some positive integer $M,$  $H_{pk}(\boldsymbol{\theta}_k)$ is a parametrized Hermitian matrix of order $2^n,$ the \textit{local Hamiltonians} for the circuit, $K$ is the total number of one-qubit and two-qubit control gates, and $P$ is the number of eigenvalues ($\neq 1$) of the unitary matrices $U_k(\boldsymbol{\theta}_k)$ corresponding to the gates. Consider a one-layer circuit given in the left side of Figure \ref{fig:PQC}, then an equivalent circuit is given in the right side, where the two-qubit control gates are separated from the strings of one-qubit gates, and hence the unitary matrix corresponding to the circuit can be written as $U(\boldsymbol{\theta})=\prod_{k=1}^K U_k(\boldsymbol{\theta}_k),$ where $U_k(\boldsymbol{\theta}_k)$ is a unitary matrix corresponding to a string of one-qubit gates or a two-qubit control gate. In Figure \ref{fig:PQC}, the unitary matrix can be decomposed into product of $8$ unitaries $U(\boldsymbol{\theta})=\prod_{k=1}^8 U_k(\boldsymbol{\theta}_k),$ for any $k,$ $U_k(\boldsymbol{\theta}_k)$ represents the unitary matrix corresponding to either a string of single-qubit gates or a two-qubit control gate with $i$-th qubit as control and $j$-th qubit as target, with either $i<j$ or $i>j.$ Then finding a (local) Hamiltonian corresponding to each $U_k(\boldsymbol{\theta}_k)$, the equation (\ref{eqn:qcdecom}) follows.

%Thus performance of an ansatz employed in a VQA can be determined by the local Hamiltonians which define the ansatz. A single layer PQC $U(\boldsymbol{\theta}^l)$ is of the form given by Figure \ref{fig:PQC}. The unitary matrices $U(\theta_{lk}),$ $1\leq k \leq K$ is a single qubit operator acting on a particular input or a control two-qubit operator. The single qubit operator is usually considered as a rotation operator $R_s(\theta_{lk})=e^{-i\theta_{lk}\sig_s},$ $s\in\{X, Y, Z\},$ and the two-qubit operator with $i$th qubit as control and $j$th qubit as target qubit can be of two types in an $n$-qubit system: $i< j$ and $i > j,$ $1\leq i,j\leq n.$ Here $\sig_s,$ $s\in\{X, Y, Z\}$ denote the Pauli matrices.

%acting on say $j$th qubit and hence $U(\theta_{lk})=$

In this paper, we consider the string of one-qubit gates comprises of arbitrary one-qubit gates, and two-qubit control gates with arbitrary  single-qubit gate for the target qubit. Then, for a string of one-qubit rotation gates it is straightforward to calculate the local Hamiltonians as described in Theorem \ref{thm:S.gate}. For a two-qubit control gate, we determine the corresponding unitary matrix $U(\boldsymbol{\theta}_k)$ of order $2^n$ with respect to the canonical basis of the $n$-qubit Hilbert space. We observe that $U(\boldsymbol{\theta}_k)$ is a $2$-sparse block diagonal matrix in general, with a symmetry in its sparsity pattern. This matrix is obtained by introducing a complete quantum binary tree of order $n,$ whose terminal nodes represent the canonical basis elements of the computational basis of the $n$-qubit Hilbert space. Due to the symmetric sparsity pattern of $U(\boldsymbol{\theta}_k),$ it is not hard to calculate the eigenpairs of it which obviously depend on the eigenpairs of the single-qubit gate that is employed on the target qubit. Consequently, we determine the local Hamiltonians $H_{kp}(\boldsymbol{\theta}_k)$ such that $$U(\boldsymbol{\theta}_k)=e^{-\iota \sum_{p=1}^P \lam_{pk}(\boldsymbol{\theta}_k)H_{pk}(\boldsymbol{\theta}_k)},$$ where $\lam_{pk}(\boldsymbol{\theta}_k)$ is a function of eigenvalues of the target single-qubit gate defined by the parameter set $\boldsymbol{\theta}_k.$ A similar method is adapted to derive local Hamiltonians corresponding to a string of arbitrary one-qubit gates in Theorem \ref{thm:sgate}. Thus we derive a local Hamiltonian decomposition of the circuit given by equation (\ref{eqn:qcdecom}), by interpreting a quantum circuit as described in Figure \ref{fig:PQC}. We provide algorithms to determine all the obtained results and demonstrate their implementation through numerical examples.  

Finally, using the obtained decomposition $U(\boldsymbol{\theta})=\prod_{k=1}^K U_k(\boldsymbol{\theta}_k)$ and exploiting the sparsity pattern of $U_k(\boldsymbol{\theta}_k),$ we determine analytical formulae of the probability amplitudes of the output quantum state for any input state for any quantum circuit defined by single-qubit and two-qubit control gates. We show that the time complexity of deriving these probability amplitudes is $O(K \, 2^n).$ Here we mention that the standard method in the literature for finding the unitary matrices corresponding to a string of one-qubit gates or a two-qubit gate in an $n$-qubit system is by finding the tensor product of the associated gates whose complexity is $O((2^n)^2)$ \cite{vidal2023paulicomposer}. Whereas, the proposed approach in this paper is optimal since there are possible $2^n$ non-zero probability amplitudes to compute for a state vector quantum simulator of any parametrized circuit on $n$-qubit systems.  We mention that the results obtained in this paper have potential applications in many areas of quantum machine learning, including generation of probability distributions which we report in a forthcoming paper. 

The remainder of the paper is organized as follows. In Section \ref{sec2}, we derive the Hamiltonians corresponding to a single-qubit and two-qubit gates in an $n$-qubit PQC and hence a local Hamiltonian decomposition of a PQC is obtained. In Section \ref{sec3}, we provide algorithms for finding probability amplitudes of any input quantum state of a PQC defined by one-qubit and two-qubit control gates. 

%In this paper, we develop a theory for determining a set of local Hamiltonians for a given PQC. Observe that unitary matrices $U(\theta_{lk}),$ $1\leq k\leq K$ are of order $2^n$ for an $n$-qubit PQC. The complexity of finding the generators ultimately boils down to the mathematical problem of finding a Hermitian matrix $H$ for a unitary matrix $U$ such that $U=e^{iH}.$ First, we determine a Hermitian matrix $H(\theta_{lk_1}, \theta_{lk_2}, \hdots, \theta_{lk_m})$ for a sequence of $m$ single qubit rotation gates such that $U(\theta_{lk_1})U(\theta_{lk_2})\hdots U(\theta_{lk_m})=e^{i H(\theta_{lk_1}, \theta_{lk_2}, \hdots, \theta_{lk_m})}.$  Then we derive explicit expression of two-qubit control-$U$ gates with $i$th qubit as control and the a single unitary matrix $U$ applied on the $j$th qubit, the target qubit $1\leq i,j\leq n$. The control-$U$ matrices are obtained by developing a notion of quantum binary tree of order $n$ whose terminal nodes represent ordered basis elements of complex Hilbert space of dimension $2^n.$ We indeed show that these controlled unitary matrices have block-diagonal form and the size of the blocks depend on whether $i < j$ or $i > j.$ Then we determine a Hermitian matrix $H$ for a control unitary $C_U$ such that $C_U=e^{-iH}$, and $H$ is obtained by utilizing eigenvectors of the single qubit operator $U.$ Finally, we provide algorithms to determine all the obtained results and demonstrate their implementation through numerical examples.  

\section{Local Hamiltonian decomposition of quantum circuits}\label{sec2}

In this section, we derive the unitary matrices corresponding to a one-qubit and two-qubit control gates and their Hamiltonians for $n$-qubit systems, $n\geq 2$.  
We introduce the notion of quantum complete binary tree which provides a useful tool for working with the standard ordered basis elements of ${\C^2}^{\otimes n},$ the $n$-times tensor product of $\C^2.$

We define a binary single-rooted tree of order $n$ whose nodes represent single/multi-qubit basis states of complex Hilbert spaces of dimension up to $2^n,$ and we call it complete quantum binary tree ($\QT_n$). Thus $\QT_n$ contains $2^n$ terminal nodes of order $n$ with two nodes of order $i$ stem from each node of order $i-1,$ $1\leq i\leq n.$ The edges which stem from a node of order $i-1$ are labelled as $0$ and $1$ and the corresponding nodes of order $i$ are represented by the states $\ket{x0}=\ket{x}\otimes \ket{0}$ and $\ket{x1}=\ket{x}\otimes \ket{1}$ respectively, where $\ket{x}$ denotes the state corresponding to the node of order $i-1.$ The root node is denoted by $t.$ If $i=1$ then the nodes which stem from $t$ are represented by $\ket{0}$ and $\ket{1}.$ Obviously, the terminal nodes of $\QT_n$ represent the standard ordered basis elements of the $2^n$ dimensional complex Hilbert space, say $\C^{2^n}$. We depict $\QT_n$ when $n=3$ in Figure \ref{fig:tree}. Then we have the following immediate observations about $\texttt{QT}_n$.

\begin{figure}%{$t$\textwidth}
				\centering
				\begin{tikzpicture}
				\draw [fill] (0, 0) circle [radius = .1];
				\node [below] at (0, 0) {$\,\,t$};
				\draw [fill] (-2, 1) circle [radius = .1];
			\node [left] at (-2, 1) {$\ket{0}$};
				\node [above] at (-1, 0.5) {$0$};
			\draw (-2,1) -- (0, 0);
				\draw [fill] (2, 1) circle [radius = .1];
				\node [right] at (2, 1) {$\ket{1}$};
				\node [above] at (1, 0.5) {$1$};
					\draw (2,1) -- (0, 0);
				\draw [fill] (-3, 2) circle [radius = .1];
				\node [left] at (-3, 2) {$\ket{00}$};
				\node [above] at (-2.5, 1.5) {$0$};
					\draw (-3,2) -- (-2, 1);
			\draw [fill] (-1, 2) circle [radius = .1];
			\node [right] at (-1, 2) {$\ket{01}$};
			\node [above] at (-1.5, 1.5) {$1$};
					\draw (-1,2) -- (-2, 1);
			\draw [fill] (3, 2) circle [radius = .1];
			\node [right] at (3, 2) {$\ket{11}$};
			\node [above] at (1.5, 1.5) {$0$};
					\draw (3,2) -- (2, 1);
			\draw [fill] (1, 2) circle [radius = .1];
			\node [left] at (1, 2) {$\ket{10}$};
			\node [above] at (2.5, 1.5) {$1$};
					\draw (1,2) -- (2, 1);
			\draw [fill] (-3.5, 3) circle [radius = .1];
			\node [above] at (-3.5, 3) {$\ket{000}$};
			\node [above] at (-3.5, 2.3) {$0$};
					\draw (-3.5,3) -- (-3, 2);
			\draw [fill] (-2.5, 3) circle [radius = .1];
			\node [above] at (-2.5, 3) {$\ket{001}$};
			\node [above] at (-2.5, 2.3) {$1$};
					\draw (-2.5,3) -- (-3, 2);
			\draw [fill] (-1.5, 3) circle [radius = .1];
			\node [above] at (-1.5, 3) {$\ket{010}$};
			\node [above] at (-1.5, 2.3) {$0$};
					\draw (-1.5,3) -- (-1, 2);
			\draw [fill] (-0.5, 3) circle [radius = .1];
			\node [above] at (-0.5, 3) {$\ket{011}$};
			\node [above] at (-0.5, 2.3) {$1$};
					\draw (-0.5,3) -- (-1, 2);
			\draw [fill] (0.50, 3) circle [radius = .1];
			\node [above] at (0.50, 3) {$\ket{100}$};
			\node [above] at (0.5, 2.3) {$0$};
					\draw (0.5,3) -- (1, 2);
			\draw [fill] (1.5, 3) circle [radius = .1];
			\node [above] at (1.5, 3) {$\ket{101}$};
			\node [above] at (1.5, 2.3) {$1$};
					\draw (1.5,3) -- (1, 2);
			\draw [fill] (2.5, 3) circle [radius = .1];
			\node [above] at (2.5, 3) {$\ket{110}$};
			\node [above] at (2.5, 2.3) {$0$};
					\draw (2.5,3) -- (3, 2);
			\draw [fill] (3.5, 3) circle [radius = .1];
			\node [above] at (3.5, 3) {$\ket{111}$};
			\node [above] at (3.5, 2.3) {$1$};
					\draw (3.5,3) -- (3, 2);
			\end{tikzpicture}
				\caption{$\texttt{QT}_3$}
				\label{fig:tree}
			\end{figure}

\begin{enumerate}
    \item[(a)] Any node of order $i\geq 1$ in $\QT_n$ represents an $i$-qubit basis state. The number of nodes of order $i$ that represent $i$-qubit states with last qubit $\ket{1}$ is $2^{i-1}$ which equals the number of nodes of order $i-1.$
    \item[(b)] The number of $n$-qubit basis states that originate from a node of order $i$ in $\QT_n$ is $2^{n-i.}$ The nodes of order $i$ in $\QT_n$ whose last qubit is $\ket{1}$ give rise to $n$-qubit states that correspond to the terminal nodes of $\QT_n,$ are those whose $i$-th qubit is $\ket{1}.$ These $n$-qubit states correspond to the ordered basis elements $\ket{xs}=\ket{x}\otimes \ket{s}$ where $\ket{s}$ belongs to the ordered basis of the complex Hilbert space of dimension $n-i$ and $\ket{x}$ is the basis state corresponding to the node of order $i,$ the $i$-qubit state.
%\item[(c)] 
\end{enumerate}

Now we have the following proposition which will be used in the sequel. 

\begin{proposition}\label{prop:eigexp}
Let $X\in\C^{n\times n}$ be a Hermitian matrix. For $\lam\in\R,$ $0\neq x\in\C^n,$ the pair $(\lam, x)$ is an eigenpair of $X$ if and only if $(e^{-i\lam},x)$ is an eigenpair of $Y=e^{-iX}.$  
\end{proposition}
\pf Let $(\lam_j,x_j),$ $j=1,\hdots,n$ be the set of orthonormal eigenpairs of $X.$ Then $X=\widetilde{X}D_\lam\widetilde{X}^\dagger$ where $\widetilde{X}=\bmatrix{x_1&\hdots&x_n}$ and $D_\lam=\diag(\lam_1,\hdots,\lam_n).$ Now $$Y=e^{-i\widetilde{X}D_\lam\widetilde{X}^\dagger}=\widetilde{X}e^{-iD_\lam}\widetilde{X}^\dagger=\sum_{j=1}^n e^{-i\lam_j}x_jx_j^\dagger.$$ Then it can be easily checked that $Yx_j=e^{-i\lam_j}x_j.$ The converse of the theorem follows similarly. $\hfill{\square}$

We denote the computational basis of $\C^{2^n}$ as $\{\ket{k_1k_2\hdots k_n}: k_j\in\{0,1\}, j=1,\hdots, n\}.$ Then the $k$-th basis element is given by $\ket{k_1k_2\hdots k_n}$, where $k=\sum_{\alpha=1}^n k_\alpha 2^{n-\alpha},$ $k_\alpha\in\{0,1\},$ $0\leq k\leq 2^{n-1}.$ Observe that the terminal nodes of $\QT_n$ represent the elements of the ordered canonical basis from left to right. The computational basis of $1$-qubit system is formed by $\ket{0}=\bmatrix{1\\ 0}$ and $\ket{1}=\bmatrix{0\\ 1}.$ 

First we derive generators of two-qubit control gates. Note that a fundamental difference between one-qubit and two-qubit control gates is that control gates are responsible for generating entanglements between qubits in the circuit and hence circuits consisting of sequence of two-qubit control gates are known as entanglers.

\subsection{Two-qubit control gates}

In what follows, we consider finding the unitary matrix $C_U\in\C^{2^n\times 2^n}$ corresponding to a two-qubit control-$U$ gate with $U\in\C^{2\times 2}$ as the single-qubit gate, and a Hermitian matrix $H$ such that $C_U=e^{-\iota H}$. We assume that the underlying system is an $n$-qubit system, and the control and target qubits are the $i$-th and $j$-th qubits respectively, $1\leq i,j\leq n.$ Thus the gate $U$ applied on the $j$-th qubit only when the state of the $i$th qubit is $\ket{1}.$ We deal with $i<j$ and $i>j$ separately as follows.

\subsubsection{$i < j$}
 We first determine $C_U$ corresponding to the computational (canonical) ordered basis of $\C^{2^n}$ and then derive symbolic representations of its eigenpairs defined by the symbolic eigenpairs of $U.$ Then, by employing Proposition \ref{prop:eigexp}, a Hermitian matrix $H$ of order $2^n$ is obtained such that $C_U=e^{-\iota H}.$ 

Let $U=\bmatrix{u_{11}&u_{12}\\ u_{21}&u_{22}}.$ Then for any basis element $\ket{k_1\hdots k_i\hdots k_j\hdots k_n}$ of $\C^{2^n},$  \begin{equation}\label{cu1}C_U\ket{k_1\hdots k_i\hdots k_j\hdots k_n}=\left\{
  \begin{array}{ll}
u_{11}\ket{k_1\hdots k_i\hdots k_{j-1} 0k_{j+1}\hdots k_n} + u_{21}\ket{k_1\hdots k_i\hdots k_{j-1} 1k_{j+1}\hdots k_n}, & \\   \hfill{\hbox{\mbox{if}\,\, $\ket{k_j}=\ket{0}$}} \\
u_{12}\ket{k_1\hdots k_i\hdots k_{j-1} 0k_{j+1}\hdots k_n} + u_{22}\ket{k_1\hdots k_i\hdots k_{j-1} 1k_{j+1}\hdots k_n}, & \\  \hfill{\hbox{\mbox{if}\,\, $\ket{k_j}=\ket{1}$}}
  \end{array}
\right.\end{equation} if $\ket{k_i}=\ket{1};$ and $C_U$ acts as an identity operator when $\ket{k_i}=\ket{0}$. 

Note that if the input $n$-qubit state $\ket{k_1\hdots k_i\hdots k_j\hdots k_n}$ is the $l$-th basis element in the standard ordered  basis of $2^n$ dimensional Hilbert space then the corresponding output under $C_U$ can be written as a linear combination of $l$-th and $(l+2^{n-j})$-th basis elements if $\ket{k_j}=\ket{0},$ whereas the output is a linear sum of $l$-th and $(l-2^{n-j})$-th basis states if $\ket{k_j}=\ket{1}$ (note that if $k_j=1$ then $l> 2^{n-j}$), $1\leq l\leq 2^n.$

As discussed above, the number of $n$-qubit basis states with control qubit $\ket{k_i}=\ket{1}$ equals the number of nodes of order $i-1$ in $\QT_n$ i.e. $2^{i-1}.$ Further, the two nodes of order $i$ that stem from each node of order $i-1$ with edge labels $0$ and $1$ correspond to $i$-qubit states with $i$-th qubit $\ket{0}$ and $\ket{1}$ respectively and $C_U$ acts nontrivially on the $n$-qubits that are stemmed from the later. Since there are $2^{n-i}$ terminal nodes that are originated from every node of order $i$ in $\QT_n,$ $C_U$ acts nontrivially on every alternative collection of consecutive $2^{n-i}$ $n$-qubit basis states starting from $(2^{n-i}+1)$-th $n$-qubit basis state in the standard ordered basis of $\C^{2^n}.$  

Let $B_l, l=2\beta,$ $\beta\in\{1,\hdots,2^{i-1}\}$ denote the ordered sets of consecutive ordered $2^{n-i}$ basis states that corresponds to terminal nodes of $\QT_n$ and $C_U$ acts non-trivially on each element of $B_l$, each of which is originated from a node of order $i$ which corresponds to an $i$-qubit state with the last qubit $\ket{1}.$ Besides, from equation (\ref{cu1}) it follows that output of each $n$-qubit state with $i$-th qubit $\ket{1}$ i.e. for an element in $B_l,$ under $C_U$ can be written as a linear combination of itself and the state that places $2^{n-j}$ after it in the standard order of the basis when the $j$-th qubit is $\ket{0},$ and if the $j$-th qubit is $\ket{1}$ then the output is a linear combination of itself and a basis state which is placed before $2^{n-j}$ in the standard order of the canonical basis. This process repeats for each set $B_l$ and we need to construct the matrix representation of $C_U.$ Its action is non-trivial on the ordered basis elements of $\C^{2^n}$ when it is restricted to the elements of $B_l$ and otherwise it acts as the identity operator. 

Thus $C_U=\diag\{U_l\in\C^{2^{n-i}\times 2^{n-i}} : 1\leq l\leq 2^{i}\}$ and $U_l=I_{2^{n-i}}$ when $l=1+2\alpha,$ $\alpha\in \{0,1,\hdots, 2^{i-1}-1\}.$ Next, for $l=2\beta,$ $\beta\in\{1,2\hdots,2^{i-1}\}$ the action of $U_l$ on the elements in $B_l$ should be written as linear combination of elements of $B_l$ following equation (\ref{cu1}) since $i<j$ i.e. $2^{n-j}<2^{n-i}$, for each $l.$ Thus $U_{2\beta}=\widehat{U}\in \C^{2^{n-i}\times 2^{n-i}}$, a fixed matrix for all $\beta.$ Finally, $C_U$ takes the diagonal block matrix form as \begin{equation}\label{cum}C_U= \diag\{\underbrace{I_{2^{n-i}}, \widehat{U}, I_{2^{n-i}},\widehat{U},\hdots,I_{2^{n-i}}, \widehat{U}}_{2^i\,\mbox{-blocks}} \}.\end{equation}

Now based on the above observations, we determine a Hamiltonian  $H$ corresponding to the unitary operator $C_U$ such that $C_U=e^{-\iota H}.$ First we state the following lemma which can be proved easily.

\begin{lemma}\label{lem:ep}
Let $A=\diag\{A_1,A_2,\hdots,A_k\}$ be a diagonal block matrix with $A_l\in\C^{m\times m},$ $1\leq l\leq k.$ If $(\lam^{(l)},v^{(l)})$ is an eigenpair of $A_l$ for some $l$ then $(\lam^{(l)}, e_l\otimes v^{(l)})$ is an eigenpair of $A,$ where $e_l$ is the $l$-th element of the standard ordered basis of $\C^k.$
\end{lemma}
%\pf The proof is obvious.

\begin{figure}[!ht]
 \centering
    {\includegraphics[width=0.80\textwidth]{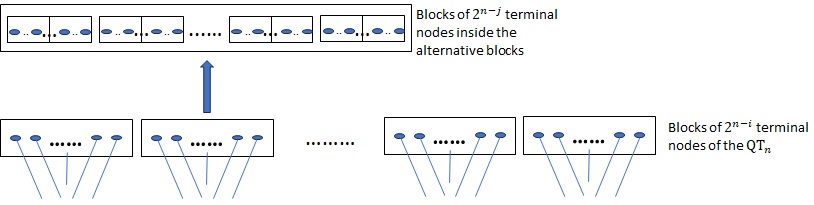} }%
 \centering \caption{Formation of the matrix $\widehat{U}^{(1)}$}
 \label{Uform1}
\end{figure}

\begin{theorem}\label{thm:ilj}
Let $C_U$ be a control-$U$ gate on a $n$-qubit system with $i$-th qubit as the control qubit and $j$-th qubit as target qubit, $1\leq i<j\leq n.$ Then $C_U$ takes the form given by equation (\ref{cum}) where $\widehat{U}=\diag\{\underbrace{\widehat{U}^{(1)}, \hdots, \widehat{U}^{(1)}}_{2^{j-i-1}\,\mbox{-blocks}}\}$ and $\widehat{U}^{(1)}\in\C^{2^{n-j+1}\times 2^{n-j+1}}$ is given by equation (\ref{eqn:upq}). A Hermitian matrix $H\in\C^{2^n\times 2^n}$ such that $C_U=e^{-\iota H}$ is given by equation (\ref{eqn:H1}).
\end{theorem}

\pf Note that $C_U$ acts non-trivially on an $n$-qubit input $\ket{k_1\hdots k_i\hdots k_j\hdots k_n}$ if $\ket{k_i}=\ket{1},$ otherwise the input remains invariant. 
%It follows from $\QT_n$ that number of $n$-qubit basis states with control qybut $\ket{1}$ equals the number of nodes of order $i-1$ in $\QT_n.$ Further, the two nodes of order $i$ that stem from each node of order $i-1$ with edge labels $0$ and $1$ correspond to $i$-qubit states with $i$-th qubit $\ket{0}$ and $\ket{1}$ respectively and $C_U$ acts nontrivially on the $n$-qubits that are stemmed from the later. Since there are $2^{n-i}$ terminal nodes that are stemmed from any node of order $i$ in $\QT_n,$ $C_U$ acts nontrivially on every alternative collection of consecutive $2^{n-i}$ $n$-qubit states starting from $(2^{n-i}+1)$-th $n$-qubit basis state. Hence, $C_U=\diag\{U_l\in\C^{2^{n-i}}\times 2^{n-i} : 1\leq l\leq 2^{i}\}$ where $U_l=I_{2^{n-i}}$ if $l=1+2{\alpha},$ $\alpha=0,1,\hdots,2^{i-1}-1,$ and otherwise if $l=2\beta,$ $\beta=1,\hdots, 2^{i-1}$ then $U_l$ can be derived as follows.
A basis element  $\ket{k_1\hdots k_i\hdots k_j\hdots k_n}$ corresponds to the non-negative integer $k=\sum_{\alpha=1}^n k_\alpha 2^{n-\alpha},$  $k_\alpha\in\{0,1\}.$ Thus the ordered basis of the $2^n$ dimensional Hilbert space can be written as $\{\ket{k} : 0\leq k\leq 2^{n-1}\}$ with the one-one correspondence $\ket{k} \leftrightarrow \ket{k_1\hdots k_i\hdots k_j\hdots k_n}.$ Then from equation (\ref{cu1}) we obtain \begin{equation}
    \bra{k}C_U\ket{k} = u_{11}, \,\,
    \bra{k}C_U\ket{k+2^{n-j}} = u_{12},\,\,  
    \bra{k+2^{n-j}}C_U\ket{k} = u_{21}, \,\, 
    \bra{k+2^{n-j}}C_U\ket{k+2^{n-j}} = u_{22} 
\end{equation} when $\ket{k_i}=\ket{1}.$ Further, $2^{n-j}<2^{n-i}$ since $i<j,$ and hence $C_U$ takes the form as mentioned in equation (\ref{cum}).

%Let $B_i$ denote the ordered set of consecutive ordered $2^{n-i}$ basis states that corresponds to terminal nodes of $\QT_n$ and $C_U$ acts on them non-trivially, each of which is originated from a node of order $i$ which corresponds to an $i$-qubit state with the last state $\ket{1}.$ Besides, from equation (\ref{cu}) it follows that output of each $n$-qubit state with $i$-th qubit $\ket{1}$ under $C_U$ can be written as a linear combination of itself and the state that places $2^{n-j}$ after it in the standard order of the basis when the $j$-th qubit of is $\ket{0},$ and if the $j$-th qubit is $\ket{1}$ the the output is a linear combination of itself and a basis state which is placed before $2^{n-j}$ in the standard order of the basis. This process repeats for each set $B_i$ and we need to construct the matrix representation of $C_U$ when its action is restriced to the basis elements of $B_i,$ that its, $U_l, l=2\beta.$ However, it can further be observed that this  

Observe that $\widehat{U}$ acts similarly in each set $B_l,$ and the matrix representation of the action of $\widehat{U}$ on the first $2\times 2^{n-j}=2^{n-j+1}$ elements of $B_l,$  is enough for the construction of $\widehat{U}$ since $\widehat{U}$ acts similarly periodically on the remaining states in $B_l.$ The number of blocks of basis elements of $B_l$ on which $\widehat{U}$ acts on identical fashion is $2^{n-i}/(2\times 2^{n-j})=2^{j-i-1}.$ 
%$2^{n-i}/(1+2^{n-j}+2^{n-j}-1)=2^{n-i}/2^{n-j+1}=2^{j-i-1}.$ 
Therefore, $\widehat{U}$ is a block diagonal matrix with repeating diagonal blocks which are unitary matrices of order $2^{n-j+1}$ (see Figure \ref{Uform1}). Then  
\begin{equation}\label{eqn:uhatc1}\widehat{U}=\diag\{\underbrace{\widehat{U}^{(1)}, \hdots, \widehat{U}^{(1)}}_{2^{j-i-1}\,\mbox{-blocks}}\}\end{equation} with  
$\widehat{U}^{(1)}=\bmatrix{u_{pq}^{(1)}}\in \C^{2^{n-j+1} \times 2^{n-j+1}}$ and
\begin{equation}\label{eqn:upq}
u_{pq}^{(1)}=\left\{
  \begin{array}{ll}
u_{11} & \hbox{$1\leq p=q \leq 2^{n-j}$} \\
u_{12} & \hbox{$p=r, q=r+2^{n-j}, r\in\{1,\hdots,2^{n-j}\}$}\\
u_{21} & \hbox{$p=r+2^{n-j}, q=r, r\in \{1,\hdots,2^{n-j}\}$}\\
u_{22} & \hbox{$1+2^{n-j}\leq p= q\leq 2^{n-j+1}$}\\
0 & \hbox{otherwise.}
\end{array}
\right.\end{equation} Thus $\widehat{U}^{(1)}$ is a two-sparse matrix with the following structure $$\widehat{U}^{(1)}=\bmatrix{u_{11}&&&&u_{12}&&& \\ & u_{11}&&&&u_{12}&& \\ &&\ddots&&&&\ddots& \\
&&&u_{11}&&&&u_{12} \\ u_{21}&&&&u_{22}&&& \\ & u_{21}&&&&u_{22}&& \\ &&\ddots&&&&\ddots& \\
&&&u_{21}&&&&u_{22}}.$$
Consequently, the expression of $\widehat{U}$ follows.
This concludes the proof of the first part of the theorem about the construction of $C_U$.

Next we consider finding a Hermitian matrix $H$ such that $C_U=e^{-\iota H}.$ Using Proposition \ref{prop:eigexp}, it is enough to determine a complete set of eigenpairs of $C_U$ that enables to derive eigendecomposition of $H$ and consequently an explicit formula for $H$ can be obtained. First observe that the one-qubit gate $U$ is embedded in $\widehat{U}^{(1)}\in \C^{2^{n-j+1}\times 2^{n-j+1}}$ as submatrix corresponding to the indices $\{r,r+2^{n-j}\}\times \{r,r+2^{n-j}\}$ for any $r\in \{1,\hdots,2^{n-j}\}.$ Let $\ket{u_1}=\bmatrix{\widehat{u}_{11} & \widehat{u}_{12}}^T$ and $\ket{u_2}=\bmatrix{\widehat{u}_{21} & \widehat{u}_{22}}^T$ be the orthonormal eigenvectors of $U$ corresponding to the eigenvalues $\lam_1$ and $\lam_2$ respectively. Then for any $r,$ the vector $\ket{v_r^{(1)}}_s=[v_{s,r}]_{l=1}^{2^{n-j+1}}\in \C^{2^{n-j+1}}$ given by 
\begin{equation}\label{eqn:upq1}
[v_{s,r}]_l=\left\{
  \begin{array}{ll}
\widehat{u}_{s1} & \hbox{if $l=r$} \\
\widehat{u}_{s2} & \hbox{if $l=r+2^{n-j}$}\\
0 & \hbox{otherwise.}
\end{array}
\right.\end{equation} is an eigenvector of $\widehat{U}^{(1)}$ corresponding to the eigenvalue $\lam_s, s=1,2.$ Thus $\left(\lam_s,\ket{v_r^{(1)}}_s\right), s=1,2,$  $r\in\{1,\hdots,2^{n-j}\}$ provide a complete list of eigenpairs of $\widehat{U}^{(1)}.$  

Now since $\widehat{U}$ is a diagonal block matrix with $2^{j-i-1}$ blocks $\widehat{U}^{(1)},$ by Lemma \ref{lem:ep} a complete list of orthonormal eigenvectors of $\widehat{U}$ are given by $$\left(\lam_s, \ket{v_{l,s,r}}=\ket{e_l}\otimes \ket{v_r^{(1)}}_s\right)$$ for any $\ket{e_l}$ belongs to the standard ordered basis of $\C^{2^{j-i-1}}$, $1\leq l\leq 2^{j-i-1}$ $s=1,2$, and $r\in \{1,\hdots,2^{n-j}\}.$ Note that eigenvalues of $U$ are eigenvalues of $\widehat{U}$ and each eigenvalue of $U$ has algebraic multiplicity $2^{n-i-1}$ as an eigenvalue of $\widehat{U}.$

Further, since $C_U=\diag\{\underbrace{I_{2^{n-i}}, \widehat{U}, I_{2^{n-i}},\widehat{U},\hdots,I_{2^{n-i}}, \widehat{U}}_{2^i\,\mbox{-blocks}} \},$ by Lemma \ref{lem:ep} it follows that the complete list of eigenpairs of $C_U$ is $$(1,\ket{f_{m_1}}\otimes \ket{e_{m_2}} ), (\lam_s,\ket{f_{m_3}}\otimes \ket{v_{l,s,r}})$$ where $\{\ket{f_{m_1}} \in\C^{2^i}: m_1=1+2\alpha, \alpha=0,1,\hdots,2^{i-1}-1 \}$ is the subset of the standard ordered basis $F_m=\{\ket{f_m} : 1\leq m\leq 2^i\}$ of $\C^{2^i}$, $\{\ket{e_{m_2}} : 1\leq m_2 \leq 2^{n-i}\}$ is the standard ordered basis of $\C^{2^{n-i}},$ $\{\ket{f_{m_3}} : m_3=2\beta, \beta=1,2,\hdots,2^{i-1}\}$ is a subset of $F_m$, and $(\lam_s,\ket{v_{l,s,r}})$ are eigenpairs of $\widehat{U}.$ Thus the eigendecomposition of $C_U$ is given by $$C_U=\sum_{m_1,m_2} (\ket{f_{m_1}}\otimes \ket{e_{m_2}})(\bra{f_{m_1}}\otimes \bra{e_{m_2}}) + \sum_{m_3,l,s,r} \lam_s (\ket{f_{m_3}}\otimes \ket{v_{l,s,r}})(\bra{f_{m_3}}\otimes \bra{v_{l,s,r}}).$$

Finally, by Proposition \ref{prop:eigexp} observe that eigenvalues of a desired $H$ are given by $z$ which correspond to the eigenvalues $\lam$ of $C_U$ such that $e^{-\iota z}=\lam,$ and the eigenvectors of $H$ are same as the corresponding eigenvectors of $C_U.$ Since some of the eigenvalues of $C_U$ are $1,$ the corresponding eigenvalues of $H$ are $0,$ and hence they do not contribute to the construction of $H,$ the remaing eigenvalues of $H$ are $z_s$ such that $\lam_s=e^{-\iota z_s}.$ Thus we conclude that \begin{equation}\label{eqn:H1}
    H=\sum_{m_3,l,s,r} z_s \, (\ket{f_{m_3}}\bra{f_{m_3}}\otimes \ket{v_{l,s,r}}\bra{v_{l,s,r}}).
\end{equation}
This completes the proof. $\hfill{\square}$

\begin{example}\label{exp1}
Now we consider an example of $n=5$-qubit system. We assume that $i$-th qubit is control and $j$-th qubit is the target. In the following we determine explicit formula for $\widehat{U}.$
\begin{itemize}
    \item[(a)] Let $i=2$ and $j=3.$ Then $$\widehat{U}=\bmatrix{u_{11} & 0&0&0&u_{12}&0&0&0\\ 0 & u_{11} & 0& 0& 0& u_{12} &0&0\\ 0 & 0 & u_{11} & 0&0&0& u_{12} &0\\ 0 & 0&0& u_{11} &0&0&0& u_{12} \\ u_{21}&0&0&0&u_{22}&0&0&0 \\ 0& u_{21} &0&0&0& u_{22} &0&0 \\ 0&0& u_{21} &0&0&0& u_{22} &0 \\ 0&0&0& u_{21} &0&0&0& u_{22}}$$ and consequently $C_U=\diag\{I_{2^3},\widehat{U},I_{2^3},\widehat{U}\}.$

\item[(b)] Let $i=2$ and $j=4.$ Then $\widehat{U}=\diag\{\widehat{U}^{(1)},\widehat{U}^{(1)}\}$ where $$\widehat{U}^{(1)}=\bmatrix{u_{11}&0&u_{12}&0 \\ 0&u_{11}&0&u_{12}\\ u_{21}&0&u_{22}&0 \\ 0&u_{21}&0&u_{22}}$$ and hence $C_U=\diag\{I_{2^3},\widehat{U}^{(1)},\widehat{U}^{(1)},I_{2^3},\widehat{U}^{(1)},\widehat{U}^{(1)}\}.$ 
\end{itemize}
\end{example}

Now we describe algorithms based on the proof of Theorem \ref{thm:ilj} for generation of the unitary matrix $C_U$ and the Hermitian matrix $H$ such that $C_U=e^{-\iota H}.$ Algorithm \ref{alg:widehat1_u} describes construction of $\widehat{U}^{(1)}$, which is then used in Algorithm \ref{alg:MatRep_CU} for the construction of $C_U.$ Next, Algorithm \ref{alg:eigvec_widehat1_u} delivers symbolic expression of eigenpairs of $\widehat{U}^{(1)},$ which is further used to derive a Hermition matrix $H$ in Algorithm \ref{alg:oneConnMatrix} such that $C_U=e^{-\iota H}.$

\begin{algorithm}
    \caption{Construction of $\widehat{U}^{(1)}$}
    \label{alg:widehat1_u}
    \begin{algorithmic}[1]
      	\Require The fixed variables $n,i,j,$ and $1\leq i<j\leq n.$ A unitary matrix $U=\bmatrix{u_{11} & u_{12} \\ u_{21} & u_{22}}.$  
		\Ensure Generation of the symbolic unitary matrix $\widehat{U}^{(1)}=\left[u_{pq}^{(1)}\right]\in \C^{2^{n-j+1}\times 2^{n-j+1}}$.
        \Statex
        \Procedure{Uhat1}{$U$}
            %\State $\mathcal{C} \gets $\Call{Cycles}{$\calD$}
            %\State $\mathcal{S} \gets $ all pairwise vertex disjoint subsets of $\mathcal{C}$
            \State $\widehat{U}^{(1)} \gets 0$
            \ForAll{$1 \leq p,q \leq 2^{n-j}$}
            \For{$r\gets 1$ to $2^{n-j}$}
	\If{$p=r$ \, and \, $q=r+2^{n-j}$}
	\State $u^{(1)}_{pq}\gets u_{12} $
	\EndIf
	%\For{$r\gets 1$ to $2^{n-j}$}
	\If{$q=r$ \, and \, $p=r+2^{n-j}$}
	\State $u^{(1)}_{pq}\gets u_{21} $
	\EndIf
	\EndFor
	  	\If{$p=q$ \, and \, $1\leq p\leq 2^{n-j}$}
	\State $u^{(1)}_{pq} \gets u_{11}$
	\EndIf
    	\If{$p=q$ \, and \, $1+2^{n-j}\leq p\leq 2^{n-j+1}$}
	\State $u^{(1)}_{pq} \gets u_{22}$
	\EndIf
	%{$p=q$ $\&\& p\geq 1+2^{n-j} || p\leq 2^{n-j+1}$ }
%	\State $u^{(1)}[p][q] \gets 0$
	%\Endelse
	\EndFor
    \State \Return $\widehat{U}^{(1)}$
        \EndProcedure
    \end{algorithmic}
\end{algorithm}

\begin{algorithm}
    \caption{Matrix representation of $C_U$}
    \label{alg:MatRep_CU}
    \begin{algorithmic}[1]
      	\Require The fixed variables $n,i,j,$ and $1\leq i<j\leq n.$ A unitary matrix $U=\bmatrix{u_{11} & u_{12} \\ u_{21} & u_{22}}.$  
		\Ensure The matrix representation of the control two-qubit gate $C_U$ with $i$th qubit as control and $j$th qubit as target in an $n$-qubit system, $(i< j)$
        \Statex
        \Procedure{MRep}{$C_U$}
            %\State $\mathcal{C} \gets $\Call{Cycles}{$\calD$}
            %\State $\mathcal{S} \gets $ all pairwise vertex disjoint subsets of $\mathcal{C}$
            \State $M \gets \Call{Uhat1}{U}$
           \State $\widehat{U}\gets \diag\{\underbrace{M,\hdots,M}_{2^{j-i-1}-\mbox{copies}}\}$
    \State \Return $C_U=\diag\{\underbrace{I_{2^{n-i}}, \widehat{U}, I_{2^{n-i}}, \widehat{U}, \hdots, I_{2^{n-i}}, \widehat{U}}_{2^{j}\mbox{- blocks}}\}$
        \EndProcedure
    \end{algorithmic}
\end{algorithm}

\begin{algorithm}
    \caption{Computation of eigenpairs of $\widehat{U}^{(1)}$}
    \label{alg:eigvec_widehat1_u}
    \begin{algorithmic}[1]
      	\Require The fixed variables $n,i,j,$ and $1\leq i<j\leq n.$ Eigenairs $\left(\lam_s,\bmatrix{\widehat{u}_{s1} \\ \widehat{u}_{s2}}\right)$, $s=1,2$ of the one-qubit gate $U.$
		\Ensure Construction of eigenpairs $\left(\lam_s, v_{r,s}=[v_{r,s}]_{l=1}^{2^{n-j+1}}\right),$ $s=1,2$ and $r=1,\hdots, 2^{n-j}.$ 
        \Statex
        \Procedure{EigVec}{$\widehat{U}^{(1)}$}
            %\State $\mathcal{C} \gets $\Call{Cycles}{$\calD$}
            %\State $\mathcal{S} \gets $ all pairwise vertex disjoint subsets of $\mathcal{C}$
            \For{$s \gets 1$ to $2$}
            \For{$r \gets 1$ to $2^{n-j}$}
            \State $v_{r,s} \gets 0$
            \For{$l \gets 1$ to $2^{n-j+1}$}
            \If{$l=r$}
            \State $[v_{r,s}]_l\gets \widehat{u}_{s1}$
            \EndIf
            \If{$l=r+2^{n-j}$}
             \State $[v_{r,s}]_l\gets \widehat{u}_{s2}$
            \EndIf
            \EndFor
            \EndFor
            \EndFor
           \State \Return $v_{r,s}$
        \EndProcedure
    \end{algorithmic}
\end{algorithm}

\begin{algorithm}
    \caption{Finding a Hamiltonian $H$ for the two-qubit control gate $C_U.$ }
    \label{alg:oneConnMatrix}
    \begin{algorithmic}[1]
      	\Require The fixed variables $n,i,j,$ and $1\leq i<j\leq n.$ Standard ordered bases $\{e_l : 1\leq l\leq 2^{j-i-1}\},$ $\{f_m : 1\leq m\leq 2^{i}\}$ of $\C^{2^{j-i-1}}$ and $\C^{2^i}$ respectively. The eigenvalues $\lam_s=e^{-\iota z_s}, s=1,2$ of $U.$ 
		\Ensure Construction of a Hermitian matrix $H$ such that $C_U=e^{-\iota H}$
        \Statex
        \Procedure{Hamiltonian}{$C_U$}
            %\State $\mathcal{C} \gets $\Call{Cycles}{$\calD$}
            %\State $\mathcal{S} \gets $ all pairwise vertex disjoint subsets of $\mathcal{C}$
            \For{$s\gets 1$ to $2$}
            \For{$r \gets 1$ to $2^{n-j}$}
            \For{$l \gets $ to $2^{j-i-1}$}
            \State $v_{l,r,s} \gets e_l\otimes \Call{EigVec}{\widehat{U}^{(1)}}$
            \EndFor
            \EndFor
            \EndFor
           \State \Return $H=\sum_{\beta=1}^{2^{i-1}} z_s \, \left(f_{2\beta} \otimes v_{l,r,s}\right) \left(f_{2\beta} \otimes v_{l,r,s}\right)^T$
        \EndProcedure
    \end{algorithmic}
\end{algorithm}

%\tb{BA: For numerical computations, start with some $U$ and then compute $H$ and finally report the error $\|C_U-e^{-iH}\|_F.$}

\subsubsection{$i > j$}

Consider a control-$U$ gate $C_U$ applied to an $n$-qubit state of the form $\ket{k} =\ket{k_1\hdots k_j\hdots k_i\hdots k_n}$ where $\ket{k_l}\in\{\ket{0},\ket{1}\},$ and $i$-th qubit $\ket{k_i}$ is the control qubit and $\ket{k_j}$ is the target qubit, $1\leq j<i\leq n$. If $U=\bmatrix{u_{11}&u_{12}\\ u_{21}&u_{22}}$ then \begin{equation}\label{cu}C_U\ket{k_1\hdots k_j\hdots k_i\hdots k_n}=\left\{
  \begin{array}{ll}
u_{11}\ket{k_1\hdots  k_{j-1} 0k_{j+1} \hdots k_i \hdots k_n} + u_{21}\ket{k_1\hdots k_{j-1} 1k_{j+1} \hdots k_i \hdots k_n}, & \\   \hfill{\hbox{\mbox{if}\,\, $\ket{k_j}=\ket{0}$}} \\
u_{12}\ket{k_1\hdots k_{j-1} 0k_{j+1} \hdots k_i \hdots k_n} + u_{22}\ket{k_1\hdots k_{j-1} 1k_{j+1} \hdots k_i \hdots k_n}, & \\  \hfill{\hbox{\mbox{if}\,\, $\ket{k_j}=\ket{1}$}}
  \end{array}
\right.\end{equation} when $\ket{k_i}=\ket{1};$ and $C_U$ acts as an identity operator on other $n$-qubit basis elements. Similar to the case $i<j,$ $C_U$ acts non-trivially on the alternative collection of $2^{n-i}$ ordered vectors that correspond to terminal nodes originated from an $i$-qubit state with $i$-th qubit $\ket{1}.$ However, the fundamental difference of the action of $C_U$ in the previous case ($i<j$) and in this case is that when $C_U$ acts on a basis state $\ket{k}$ non-trivially then the output state is a linear combination of $\ket{k}$ and another basis state which is not a terminal node that has originated from the same $i$-qubit state as $\ket{k},$ since $2^{n-j} > 2^{n-i}.$  

Thus \begin{equation}\label{cu2} C_U=\diag\{\underbrace{I_{2^{n-i}}, \widehat{U}, I_{2^{n-i}}, \widehat{U}, \hdots, I_{2^{n-i}}, \widehat{U}}_{2^{j}\mbox{- blocks}}\}\end{equation} where $\widehat{U}\in \C^{(2^{n-j+1}-2^{n-i})\times (2^{n-j+1}-2^{n-i})}$ and $I_{2^{n-i}}$ both of which appear $2^{j-1}$ times alternatively in $C_U.$ Indeed, note that $C_U$ acts as identity operator on the first $2^{n-i}$ ordered basis elements of the canonical ordered basis of $\C^{2^n}$ whose elements are represented by the terminal nodes of $\QT_n$ which are originated from the $i$-th order node corresponding to the state $\ket{0}^{\otimes^{i}}.$ Then the first instance of $\widehat{U}$ appears as $C_U$ acts non-trivially on the ordered basis elements that correspond to the terminal nodes originated from the $i$-th order node corresponding to the $i$-qubit state $\ket{0}^{\otimes i-1}\otimes \ket{1}$ in $\QT_n,$ the second basis element in the ordered basis of $2^i$-dimensional Hilbert space. Then we have the following theorem.

\begin{figure}[!ht]
 \centering
    {\includegraphics[width=0.80\textwidth]{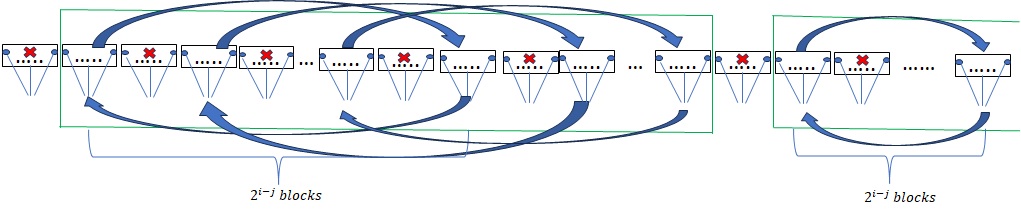} }%
 \centering \caption{Formation of the matrix $\widehat{U}$, the block in the green rectangle. The small blocks are terminal nodes of $\QT_n,$ and the red cross on alternatives of them represents that $C_U$ does not act on the basis elements inside it.}
 \label{Uform2}
\end{figure}

\begin{theorem}\label{thm:igj}
Let $C_U$ denote a control-$U$ gate on an $n$-qubit system with $i$-th qubit as the control qubit and $j$-th qubit as target qubit, $1\leq j<i\leq n.$ Then $C_U$ takes the form given by equation (\ref{cu2}) where $\widehat{U}$ is given by equations (\ref{eqn:uwide})-(\ref{eqn:dpq}). A Hermitian matrix $H\in\C^{2^n\times 2^n}$ such that $C_U=e^{-\iota H}$ is given by equation (\ref{eqn:Hamil2}).
\end{theorem}

\pf From the discussion above, $C_U$ acts non-trivially on alternative blocks of $2^{n-i}$ ordered basis states in $2^n$-dimensional Hilbert space. Each of these blocks contains $2^{n-i}$ terminal nodes in $\QT_n$ that are originated from a node of order $i.$ Let us denote the blocks as $B_l, 1\leq l\leq 2^i,$ where each $B_l$ contains $2^{n-i}$ standard ordered basis elements of $\C^{2^n}.$ The matrix $C_U$ acts nontrivially on the elements of $B_l$ for alternative $l$ (when $l$ is even) starting from $l=2.$ 

%when when Thus $C_U$ acts as identity operator for all the states belonging to $B_{2l_1+1}, 0\leq l_1 \leq 2^{i-1}-1$, and otherwise $C_U$ acts non-trivially following equation (\ref{cu}) for all the basis states belonging to $B_{2l_2}, 1\leq l_2\leq 2^{i-1}.$ 

%Thus $C_U$ is given by equation (\ref{cu2}), where $\widehat{U}$ acts on all the states in $B_{l_2}$ for all $l_2.$

Let $\ket{k}\in B_{l}$ for some even $l.$ Then $C_U\ket{k}=u_{11}\ket{k}+u_{21}\ket{k_+}$ or $C_U\ket{k}=u_{12}\ket{k_-}+u_{22}\ket{k}$ where $\ket{k_+}$ and $\ket{k_-}$ denote the state which is placed after and before $2^{n-j} >2^{n-i}$ states from the state $\ket{k}$ in the standard ordered basis of the $2^n$-dimensional Hilbert space respectively, as follows from equation (\ref{cu}). The number of blocks $B_l$ that lie in between $\ket{k}$ and $\ket{k_+}$ or $\ket{k_-}$ is $\alpha = 2^{n-j}/2^{n-i}=2^{i-j},$ and $C_U$ acts non-trivially on each such block alternatively. Thus $\widehat{U}$ is of the form \begin{equation}\label{eqn:uwide}\widehat{U}=\bmatrix{\widehat{U}^{(1)} \\ \widehat{U}^{(2)}\\ %\widehat{U}^{(3)} \\ 
%\widehat{U}^{(4)} \\ 
\vdots \\ \widehat{U}^{(2^{i-j}-1)} \\ \widehat{U}^{(2^{i-j})}\\ \hline \\ \underline{\widehat{U}}^{(1)} \\ \underline{\widehat{U}}^{(2)} \\ %\underline{\widehat{U}}^{(3)} \\ 
%\underline{\widehat{U}}^{(4)} \\ 
\vdots \\ \underline{\widehat{U}}^{(2^{i-j}-1)} },\end{equation} where $\widehat{U}^{(l)} =[u_{pq}^{(l)}]\in \C^{2^{n-i}\times (2^{n-j+1}-2^{n-i})}, \, \underline{\widehat{U}}^{(l)}=[\underline{u}_{pq}^{(l)}] \in \C^{2^{n-i}\times (2^{n-j+1}-2^{n-i})}, 1\leq l\leq 2^{i-j}-1,$ if $l$ is odd, and $\widehat{U}^{(l)}= [d_{pq}^{(l)}]\in \C^{2^{n-i}\times (2^{n-j+1}-2^{n-i})}$ and $\underline{\widehat{U}}^{(l)}=[\underline{d}_{pq}^{(l)}]\in \C^{2^{n-i}\times (2^{n-j+1}-2^{n-i})}$ if $l$ is even, with  
\begin{equation}\label{eqn:upq2}
u_{pq}^{(l)}=\left\{
  \begin{array}{ll}
u_{11}, & \hbox{if $q=p+(l-1)2^{n-i}, 1\leq p\leq 2^{n-i}$} \\
u_{12}, & \hbox{if $q=p+(l-1)2^{n-i}+2^{n-j}, 1\leq p\leq 2^{n-i}$}\\
0, & \hbox{otherwise.}
\end{array}
\right.\end{equation} 
\begin{equation}\label{eqn:lupq}
\underline{u}_{pq}^{(l)}=\left\{
  \begin{array}{ll}
u_{21}, & \hbox{if $q=p+(l-1)2^{n-i}, 1\leq p\leq 2^{n-i}$} \\
u_{22}, & \hbox{if $q=p+(l-1)2^{n-i}+2^{n-j}, 1\leq p\leq 2^{n-i}$}\\
0, & \hbox{otherwise.}
\end{array}
\right.\end{equation}

\begin{eqnarray}
   d_{pq}^{(l)} &=&\left\{
  \begin{array}{ll}
1, & \hbox{if $q=p+(l-1)2^{n-i}, 1\leq p\leq 2^{n-i}$} \\
%u_{12} & \hbox{if $q=p+2^{n-j}, (l-1)2^{n-i}\leq p\leq l2^{n-i}$}\\
0, & \hbox{otherwise.}
\end{array}
\right. \label{eqn:dpq} \\ 
\underline{d}_{pq}^{(l)} &=& \left\{
      \begin{array}{ll}
    1, & \hbox{if $q=p+(l-1)2^{n-i}+2^{n-j}, 1\leq p\leq 2^{n-i}$} \\
    0, & \hbox{otherwise.}
    \end{array}
    \right. \label{eqn:dupq}
\end{eqnarray}

Now we focus on finding eigenpairs of $\widehat{U}$ that will essentially determine eigenpairs of $C_U.$ Note that there is a symmetry in the sparsity pattern of $\widehat{U}$ due to the pairs of submatrices $\widehat{U}^{(l)}$ and $\underline{\widehat{U}}^{(l)}$ when $l$ is odd. The matrix $U$ appears as a submatrix given by $U_p^{(l)}=\bmatrix{u_{p(p+(l-1)2^{n-i})} & u_{p(p+(l-1)2^{n-i}+2^{n-j})} \\ \underline{u}_{p(p+(l-1)2^{n-i})} & \underline{u}_{p(p+(l-1)2^{n-i}+2^{n-j})}}$ corresponding to $p$th rows of $\widehat{U}^{(l)}$ and $\underline{\widehat{U}}^{(l)},$ $1\leq p\leq 2^{n-i}.$ Obviously, eigenpairs of $U$ are eigenpairs of $U^{(l)}_p.$

Now let $\left(\lam_s, \ket{u}_s=\bmatrix{\widehat{u}_{s1} \\ \widehat{u}_{s2}}\right),$ $s=1,2$ be a complete set of eigenpairs of $U.$ Then for any pair  $(p,l),$ with any odd value of $l,$ $1\leq p\leq 2^{n-i},$ $1\leq l\leq 2^{i-j}-1$ an eigenpair of $\widehat{U}$ is given by $$\left(\lam_s, \ket{v_{l,p}^{s}} =\widehat{u}_{s1}\ket{e_{p+(l-1)2^{n-i}}} + \widehat{u}_{s2}\ket{e_{p+(l-1)2^{n-i}+2^{n-j}}} \right).$$ Here $\{\ket{e_{\alpha}} : 1\leq \alpha \leq 2^{n-j+1} - 2^{n-i}\}$ is the canonical ordered basis of $\C^{{2^{n-j+1} - 2^{n-i}}}.$ Thus $\lam_s$ is an eigenvalue of $\widehat{U}$ with algebraic multiplicity $2^{n-i}\times 2^{i-j-1}=2^{n-j-1},$ where $p$ takes values $1$ to $2^{n-i}$ and the number of odd values of $l$ is $2^{i-j}/2=2^{i-j-1}.$ Thus two orthognormal eigenpairs of $U$ provide $2^{n-j-1}\times 2=2^{n-j}$ orthonormal eigenpairs of $\widehat{U}.$ The remaining eigenpairs of $\widehat{U}$ are given by even values of $l.$ Indeed, for any even value of $1\leq l\leq 2^{i-j}$, a set of ortonormal eigenvectors of $\widehat{U}$ are given by $\ket{e_{p+(l-1)2^{n-i}}}$ with repeated eigenvalue $1,$ that correspond to the rows of $\widehat{U}^{(l)}$ where $1\leq p\leq 2^{n-i}.$ The remaining orthonormal eigenvectors of $\widehat{U}$ are $\ket{e_{p+(2^{i-j}+l'-1)2^{n-i}}}$ where $l'$ takes the even values in the range $1\leq l'\leq 2^{i-j}-1,$ that correspond to the rows of $\underline{\widehat{U}}^{(l)}$ with even values of $l.$ Therefore, we finally conclude that $1$ is an eigenvalue of $\widehat{U}$ with algebraic multiplicity $2^{n-i}(2^{i-j} -1)=2^{n-j}-2^{n-i}.$

Now from equation (\ref{cu2}) we have \beano C_U &=& \diag\{\underbrace{I_{2^{n-i}}, \widehat{U}, I_{2^{n-i}}, \widehat{U}, \hdots, I_{2^{n-i}}, \widehat{U}}_{2^{j}\mbox{- blocks}}\} \\ 
&=& \diag\left\{\underbrace{\bmatrix{I_{2^{n-i}} & \\ & \widehat{U}}, \bmatrix{I_{2^{n-i}} & \\ & \widehat{U}}, \hdots, \bmatrix{I_{2^{n-i}} & \\ & \widehat{U}}}_{2^{j-1}\mbox{- blocks}}\right\},\eeano where $ \bmatrix{I_{2^{n-i}} & \\ & \widehat{U}}\in \C^{2^{n-j+1}\times 2^{n-j+1}}.$ For finding a Hermitian matrix $H$ for which $C_U=e^{-\iota H},$ it is enough to determine orthonormal eigenvectors of $C_U$ that correspond to eigenvalues that are not $1,$ employing Proposition \ref{prop:eigexp}. Then it is evident that $(\lam, \bmatrix{\ket{\textbf{0}}\\ \ket{v}}),$ where $\ket{\textbf{0}}$ is the zero vector of $\C^{2^{n-i}},$ is an eigenpair of $\bmatrix{I_{2^{n-i}} & \\ & \widehat{U}}$ if $(\lam, \ket{v})$ is an eigenpair of $\widehat{U},$ and the remaining eigenpairs of $\bmatrix{I_{2^{n-i}} & \\ & \widehat{U}}$ are $(1, \bmatrix{\ket{f_m} \\ \textbf{0}}),$ where $\{\ket{f_m} : 1\leq m\leq 2^{n-i}\}$ is the canonical basis of $\C^{2^{n-i}},$ where $\textbf{0}$ is the zero vector of $\C^{2^{n-j+1}-2^{n-i}}.$  

Finally, by Proposition \ref{prop:eigexp} following the similar arguments in the proof of Theorem \ref{thm:ilj} we conclude that \begin{equation}\label{eqn:Hamil2} H=\sum_{s,l,p,k} z_s \, \left(\ket{f_k}\otimes \bmatrix{\ket{\textbf{0}}\\ \ket{v_{l,p}^s}}\right)\left(\ket{f_k}\otimes \bmatrix{\ket{\textbf{0}} \\ \ket{v_{l,p}^s}}\right)^\star\end{equation} where $\lam_s=e^{-\iota\lam_s},$ $\star$ denotes the conjugate transpose, $\{\ket{f_k} : 1\leq k \leq 2^{j-1}\}$ is the canonical basis of $\C^{2^{j-1}},$ $\ket{\textbf{0}}\in\C^{2^{n-i}}.$ $\hfill{\square}$

\begin{example}
Consider a $5$-qubit system.
\begin{itemize}
    \item[(a)] with $i=3$ as the control qubit and $j=2$ as the target qubit. Then $C_U=\diag\{I_{2^{2}}, \widehat{U}, I_{2^{2}}, \widehat{U}\},$ $\widehat{U}\in \C^{3\times 2^{2}}$ given by 
$$\widehat{U}=\bmatrix{U_{11} & 0 & U_{12} \\ 0 & I_{2^2} & 0 \\ U_{21} & 0 & U_{22}}, U_{pq}=\bmatrix{u_{pq} &&& \\ & u_{pq}&& \\ && u_{pq}& \\ &&& u_{pq}}, 1\leq p,q\leq 2.$$
%\bmatrix{  u_{11} & 0 & 0 & 0 & 0 & 0 & 0 & 0 & u_{12} & 0 & 0 & 0 \\  0 & u_{11} & 0 & 0 & 0 & 0 & 0 & 0 & 0 & u_{12} & 0 & 0 \\    0 & 0 & u_{11} & 0 & 0 & 0 & 0 & 0 & 0 & 0 & u_{12} & 0  \\ 0 & 0 & 0 & u_{11} & 0 & 0 & 0 & 0 & 0 & 0 & 0 & u_{12} \\ 0 & 0 & 0 & 0 & 1 & 0 & 0 & 0 & 0 & 0 & 0 & 0 \\  0 & 0 & 0 & 0 & 0 & 1 & 0 & 0 & 0 & 0 & 0 & 0 \\  0 & 0 & 0 & 0 & 0 & 0 & 1 & 0 & 0 & 0 & 0 & 0 \\  0 & 0 & 0 & 0 & 0 & 0 & 0 & 1 & 0 & 0 & 0 & 0 \\  U_{21} & 0 & 0 & 0 & 0 & 0 & 0 & 0 & u_{22} & 0 & 0 & 0 \\ 0 & u_{21} & 0 & 0 & 0 & 0 & 0 & 0 & 0 & u_{22} & 0 & 0 \\ 0 & 0 & u_{21} & 0 & 0 & 0 & 0 & 0 & 0 & 0 & u_{22} & 0 \\ 0 & 0 & 0 & u_{21} & 0 & 0 & 0 & 0 & 0 & 0 & 0 & u_{22}}$$

\item[(b)] $i=4, j=2.$ Then $C_U=\diag\{I_{2}, \widehat{U}, I_{2}, \widehat{U}\},$ where $\widehat{U}\in \C^{14\times 14}$
$$\widehat{U}=\bmatrix{U_{11} & 0 & 0 & 0 & U_{12} & 0 & 0\\ 0 & I_2 & 0 & 0 & 0 &0 & 0\\ 0 & 0 &U_{11} & 0 &0 &0 & U_{12} \\ 0 & 0& 0& I_2 &0 & 0 & 0 \\ U_{21} & 0 & 0 & 0 & U_{22} & 0 & 0 \\ 0 & 0 & 0 & 0 & 0 & I_2 & 0 \\ 0 & 0 & U_{21} & 0 & 0 & 0 & U_{22}}, U_{pq}=\bmatrix{u_{pq}& \\ & u_{pq}}, 1\leq p,q\leq 2.$$

\end{itemize}

\end{example}

\begin{example}
Let $C_U$ be a control-$U$ gate for a $2$-qubit system, where $U=\bmatrix{u_{11} & u_{12} \\ u_{21}&u_{22}}$ is a single-qubit gate. Let $\lam_s=e^{-\iota z_s},$ $s=1,2$ be the eigenvalues of $U.$
\begin{itemize}
    \item[(a)] If the first qubit is the control and second quibit is the target then the set of orthonormal eigenpairs of $C_U$ are given by $(1, \ket{00}), (1,\ket{01}),$ $(\lam_1, \ket{1}\otimes \ket{u_1})$ and $(\lam_2, \ket{1}\otimes \ket{u_2})$ where $(\lam_1,\ket{u_1})$ and $(\lam_2,\ket{u_2})$ are orthonomral eigenpairs of $U.$ Consequently, a Hamiltonian $H$ such that $C_U=e^{-\iota H}$ is given by $$H=\sum_{j=1}^2 z_j (\ket{1}\bra{1}\otimes \ket{u_j}\bra{u_j}).$$ 
    %where $\mathrm{Arg}(z)$ denotes the principal argument of $z.$ 
    \item[(b)] If the second qubit is the control and first qubit is the target then the set of orthonormal eigenpairs of $C_U$ are given by $(1, \ket{00}), (1,\ket{10}),$ $(\lam_1, \ket{\widetilde{u}_1}=\bmatrix{0 \\\widetilde{u}_{11} \\ 0\\ \widetilde{u}_{12}})$ and $(\lam_2, \ket{\widetilde{u}_2}=\bmatrix{0\\\widetilde{u}_{21} \\0\\ \widetilde{u}_{22}})$ where $(\lam_1,\ket{u_1}=\bmatrix{\widetilde{u}_{11} \\ \widetilde{u}_{12}})$ and $(\lam_2,\ket{u_2}=\bmatrix{\widetilde{u}_{21} \\ \widetilde{u}_{22}})$ are orthonomral eigenpairs of $U.$ Consequently, a Hamiltonian $H$ such that $C_U=e^{-\iota H}$ is given by $$H= z_1 \ket{\widetilde{u}_1} \bra{\widetilde{u}_1} +  z_2 \ket{\widetilde{u}_2} \bra{\widetilde{u}_2}.$$
\end{itemize}

\end{example}
\pf The proof follows by Theorem (\ref{thm:ilj}) and (\ref{thm:igj}). Indeed note that $C_U=\bmatrix{I_2 & 0\\ 0 & U}$ for ($a$) and $C_U=\bmatrix{1&0&0&0 \\ 0 &u_{11}&0&u_{12}\\ 0&0&1&0\\ 0&u_{21}&0 &u_{22}}$ for ($b$). \hfill{$\square$}

Now we describes the algorithms based on the proof of Theorem \ref{thm:igj}. First, Algorithm \ref{alg:widehatl_u} gives the construction of $\widehat{U}^{(l)}$ based on whether $l$ is odd or even, and Algorithm \ref{alg:widehatl_under_u} gives the construction of $\underline{\widehat{U}}^{(l)}$, These are employed to determine symbolic derivation of $C_U$ in Algorithm \ref{alg:MatRep_i>j}. Symbolic computation of eigenpairs of $\widehat{U}$ is given in Algorithm \ref{alg:ep_Uhat}, which is used in Algorithm \ref{alg:HamMat} to determine a Hamiltonian corresponding to $C_U.$

%The first part of the proof is straightforward just by writing $C_U=\bmatrix{I_2 & 0\\ 0 & U}.$ Next, for the second part, from Proposition \ref{prop:eigexp} the eigenvalues of $H$ for which $C_U=e^{-iH}$ are given by $\log 1=0, \log 1=0,$ $\log\lam_1=i\mathrm{Arg}(\lam_j),$ $j=1,2$ since $|\lam_1|=1=|\lam_2|$ and the corresponding orthogonal eigenvectors of $H$ are $\ket{00},$ $\ket{01},$ $\ket{1}\otimes\ket{u_j}$ respectively. Then the desired result follows from the eigendecomposition of $H.$ $\hfill{\square}$

\begin{algorithm}
    \caption{Construction of $\widehat{{U}}^{(l)}$ }
    \label{alg:widehatl_u}
    \begin{algorithmic}[1]
      	\Require The fixed variables $n,i,j,$ and $1\leq j < i\leq n.$ A unitary matrix $U=\bmatrix{u_{11} & u_{12} \\ u_{21} & u_{22}}.$  
		\Ensure Generation of $\widehat{U}^{(1)} \in \C^{2^{n-i}\times (2^{n-j+1}-2^{n-i})}$.
        \Statex
        \Procedure{$\widehat{U}^{(l)}$}{$U$}
        \State $\widehat{U}^{(l)} \gets 0$
        \If{$l \gets odd$}
            \For{$1 \le p \le 2^{n-i} $}
                \For{$1 \le q \le 2^{n-i+1}-2^{n-i}$}
                    \If{$q = p+(l-1)2^{n-i}$}
                        \State $u_{pq} \gets 1$
                    \EndIf
                \EndFor
            \EndFor
        \EndIf
        \If{$l \gets even$}
            \For{$1 \le p \le 2^{n-i} $}
                \For{$1 \le q \le 2^{n-i+1}-2^{n-i}$}
                    \If{$q = p+(l-1)2^{n-i}$}
                        \State $u_{pq} \gets u_{11}$
                    \EndIf
                    \If{$q = p+(l-1)2^{n-i}+2^{n-j}$}
                        \State $u_{pq} \gets u_{12}$
                    \EndIf
                \EndFor
            \EndFor
        \EndIf
        \State \Return  $\widehat{U}^{(l)}$
        \EndProcedure
    \end{algorithmic}
\end{algorithm}

\begin{algorithm}
    \caption{Construction of  $\underline{\widehat{U}}^{(l)}$}
    \label{alg:widehatl_under_u}
    \begin{algorithmic}[1]
      	\Require The fixed variables $n,i,j,$ and $1\leq j < i\leq n.$ A unitary matrix $U=\bmatrix{u_{11} & u_{12} \\ u_{21} & u_{22}}.$  
		\Ensure Generation of $\underline{\widehat{U}}^{(l)} \in \C^{2^{n-i}\times (2^{n-j+1}-2^{n-i})}$.
        \Statex
        \Procedure{$\underline{\widehat{U}}^{(l)}$}{$U$}
        \State $\underline{\widehat{U}}^{(l)} \gets 0$
        \If{$l \gets odd$}
            \For{$1 \le p \le 2^{n-i} $}
                \For{$1 \le q \le 2^{n-i+1}-2^{n-i}$}
                    \If{$q = p+(l-1)2^{n-i}+2^{n-j}$}
                        \State $\underline{u}_{pq} \gets 1$
                    \EndIf
                \EndFor
            \EndFor
        \EndIf
        \If{$l \gets even$}
            \For{$1 \le p \le 2^{n-i} $}
                \For{$1 \le q \le 2^{n-i+1}-2^{n-i}$}
                    \If{$q = p+(l-1)2^{n-i}$}
                        \State $\underline{u}_{pq} \gets u_{21}$
                    \EndIf
                    \If{$q = p+(l-1)2^{n-i}+2^{n-j}$}
                        \State $\underline{u}_{pq} \gets u_{22}$
                    \EndIf
                \EndFor
            \EndFor
        \EndIf
        \State \Return $\underline{\widehat{U}}^{(l)}$
        \EndProcedure
    \end{algorithmic}
\end{algorithm}

\begin{algorithm}
    \caption{Matrix representation of $C_U$}
    \label{alg:MatRep_i>j}
    \begin{algorithmic}[1]
      	\Require The fixed variables $n,i,j,$ and $1\leq j<i\leq n.$ A unitary matrix $U=\bmatrix{u_{11} & u_{12} \\ u_{21} & u_{22}}.$  
		\Ensure The matrix representation of the control two-qubit gate $C_U$ with $i$th qubit as control and $j$th qubit as target in an $n$-qubit system, $(j<i)$
        \Statex
        \Procedure{MRep}{$C_U$}
            %\State $\mathcal{C} \gets $\Call{Cycles}{$\calD$}
            %\State $\mathcal{S} \gets $ all pairwise vertex disjoint subsets of $\mathcal{C}$
            % \State $M \gets \Call{Uhat1}{U}$
           \State $\widehat{U}\gets vstack\{\widehat{U}^{1},\hdots, \widehat{U}^{2^{i-j}}, \underline{\widehat{U}}^{(1)}, \hdots, \underline{\widehat{U}}^{(2^{i-j}-1)} \}$ 
    \State \Return $C_U=\diag\{\underbrace{I_{2^{n-i}}, \widehat{U}, I_{2^{n-i}}, \widehat{U}, \hdots, I_{2^{n-i}}, \widehat{U}}_{2^{j}\mbox{- blocks}}\}$
        \EndProcedure
    \end{algorithmic}
\end{algorithm}

\begin{algorithm}
    \caption{Computation of eigenpair of $\widehat{U}$}
    \label{alg:ep_Uhat}
    \begin{algorithmic}[1]
        \Require The fixed variables $n,i,j,$ and $1\leq i>j\leq n.$Standard ordered bases $\{e_l : 1\leq l\leq 2^{n-j+1}-2^{n-i}\}$, Eigenpairs $\left(\lam_s,\bmatrix{\widehat{u}_{s1} \\ \widehat{u}_{s2}}\right)$, $s=1,2$ of the one-qubit gate $U.$
        \Ensure Construction of eigenpairs $\left(\lam_s, \ket{v_{l,p}}^{s}\right),$ $s=1,2$ , $p=1,\hdots, 2^{n-i}.$ and $l=1,\hdots, 2^{i-j}-1.$
        \Statex
        \Procedure{EigVec}{$\widehat{U}$}
        \State $\ket{v_{l,p}} \gets 0$
            \For{$s \gets 1$ to $2$}
                \For{$p \gets 1$ to $2^{n-i}$}
                    \For{$l \gets 1$ to $2^{i-j}-1$}
                        \If{$l \gets odd$}
                       \State $ \ket{v_{l,p}^{s}} \gets \widehat{u}_{s1}\ket{e_{p+(l-1)2^{n-i}}} + \widehat{u}_{s2}\ket{e_{p+(l-1)2^{n-i}+2^{n-j}}} $
                       \EndIf
                    \EndFor
                \EndFor
            \EndFor
        \State \Return $\ket{v_{l,p}}$
        \EndProcedure
    \end{algorithmic}
\end{algorithm}

\begin{algorithm}
    \caption{Finding a Hamiltonian $H$ for the two-qubit control gate $C_U.$}
    \label{alg:HamMat}
    \begin{algorithmic}[1]
      \Require The fixed variables $n,i,j,$ and $1\leq j < i\leq n.$ Standard ordered bases $\{\ket{f_{k}} : 1\leq k\leq 2^{j-1}\}$, zero vector $\{\ket{\bold{0}}: 1\leq l\leq 2^{n-j+1}-2^{n-i}\}$  The eigenvalues $\lam_s=e^{-\iota z_s}, s=1,2$ of $U.$ 
      \Ensure Construction of a Hermitian matrix $H$ such that $C_U=e^{-\iota H}$
      \Statex
      \Procedure{Hamiltonian}{$C_{U}$}
      \State $H \gets 0$
        \For{$s \gets 1$ to $2$}
            \For{$p \gets 1$ to $2^{n-i}$}
                \For{$l \gets 1$ to $2^{i-j}-1$}
                    \If{$l \gets$ odd}
                   \State $ v_{l,p,s} \gets vstack(\ket{\textbf{0}},\ket{v_{l,p}^s}) $
                    \EndIf
                \EndFor
            \EndFor
        \EndFor
      \State $H=\sum_{s,l,p,k} z_s \, \left(\ket{f_k}\otimes v_{l,p,s} \right)\left(\ket{f_k}\otimes v_{l,p,s}\right)^\star$
      \State \Return H
      \EndProcedure
    \end{algorithmic}
\end{algorithm}

\subsubsection{Numerical simulations}
% To verify the above algorithms various numerical simulations were done, where the gates were chosen from HEA Ansatz i.e. $1-2$ Controlled $R_{x}$ gate,$2-3,3-4$ Controlled $R_{x}$ gate and 100 linearly spaced values of $\theta$ were chosen between $-\pi$ to $\pi$. The error is reported as the Frobenius norm between matrices $C_{u}$ and $e^{-iH}$. From the obtained results it can be seen that the error is almost negligible and it is the same for all the three gates used.

\begin{figure}[h]
    \centering
         \subfloat [\centering ]{{\includegraphics[width=0.45\textwidth]{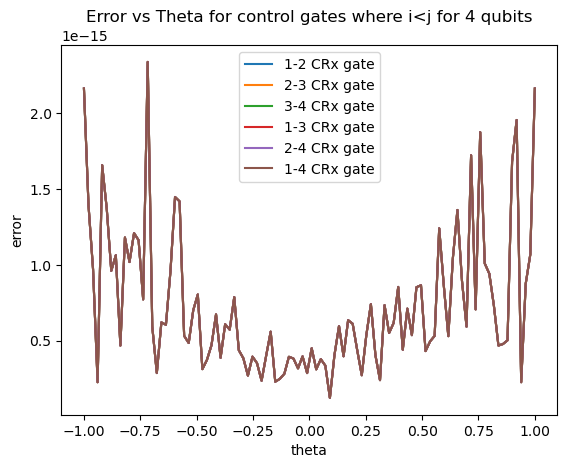} }}%
    \qquad
     \subfloat [\centering ]{{\includegraphics[width=0.45\textwidth]{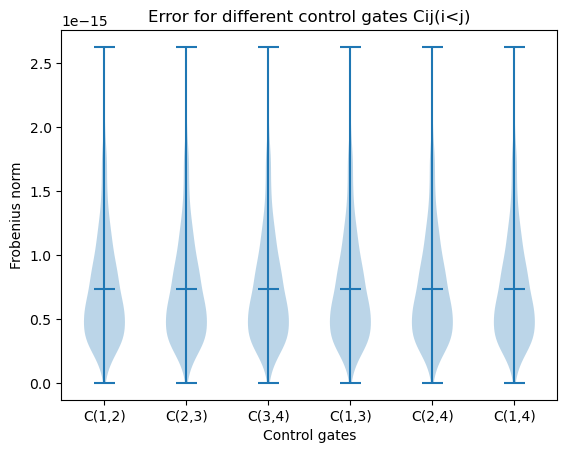} }}%
    \qquad
    \subfloat [\centering ]{{\includegraphics[width=0.45\textwidth]{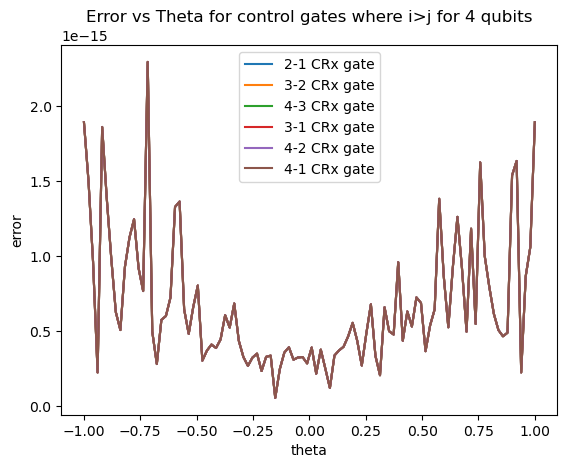} }}%
    \qquad
     \subfloat [\centering ]{{\includegraphics[width=0.45\textwidth]{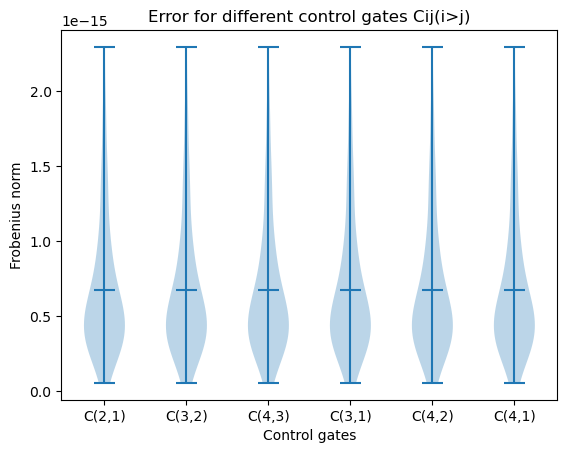} }}%
    \qquad
    \caption{(a) and (c): Errors given by $\|U(\theta) - e^{-\iota H(\theta)}\|_F$ vs $\theta$ for two-qubit control gates $i-j CR_X(\theta)$ for $100$ values of $\theta\in [-\pi, \pi]$ in a $4$-qubit system corresponding to $i<j$ and $i>j$ respectively. (b) and (d): Violin plots for the Error vs gate corresponding to (a) and (b) respectively. }
    \label{fig:numerical_sim}
\end{figure}

To verify the Algorithms \ref{alg:HamMat} and \ref{alg:oneConnMatrix} for the derivation of  various Hamiltonians associated with two-qubit control $R_X(\theta)$ ($CR_X(\theta), \theta\in [-\pi, \pi])$ gates, we assume a $4$-qubit system and calculate the error $\|U(\boldsymbol{\theta}) - e^{-\iota H(\boldsymbol{\theta})}\|_F$ through numerical simulations using the \textit{Python 3} \textit{mexp}. Here $\|\cdot\|_F$ denotes the Frobenius norm. In Fig \ref{fig:numerical_sim}, we report the errors for $100$ values of $\theta$ with $C_{ij}$ gates as the $CR_X(\theta)$ for various choices of $i$ and $j$ with $i$-th qubit as control and $j$-th qubit as target. We see that that errors are of order $10^{-15}.$

\subsection{Single-qubit gates}

\begin{figure}[!ht]
 \centering
    {\includegraphics[width=0.40\textwidth]{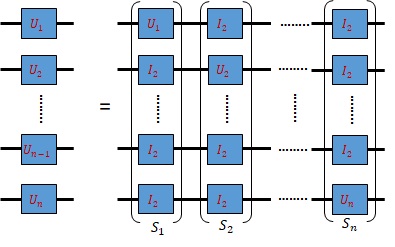} }%
 \centering \caption{Single-qubit quantum gates in PQCs}
 \label{fig:S.gate}
\end{figure}

First we consider the string of one-qubit gates in a PQC as described in Figure \ref{fig:S.gate}. In particular, in most of the common PQCs that are available in the literature, $U_j\in \{R_X, R_Y, R_Z\},$ $1\leq j\leq n.$
Here $R_X, R_Y, R_Z$ are rotation matrices for qubits around the $X,$ $Y$ and $Z$ axes respectively, on the Bloch sphere. These are given by \beano
R_X(\theta) &=& \bmatrix{\cos \frac{\theta}{2} & -\iota \sin\frac{\theta}{2} \\ -\iota \sin\frac{\theta}{2} & \cos\frac{\theta}{2}}=e^{-\iota \theta\frac{1}{2}\sigma_X}\\ 
R_Y(\theta) &=& \bmatrix{\cos \frac{\theta}{2} & -\sin\frac{\theta}{2} \\  \sin\frac{\theta}{2} & \cos\frac{\theta}{2}}=e^{-\iota \theta \frac{1}{2}\sigma_Y} \\ R_Z(\theta) &=& \bmatrix{e^{-\iota \theta/2} & 0\\ 0 & e^{\iota \theta/2}}=e^{-\iota \theta \frac{1}{2}\sigma_Z}, \eeano where $\sigma_X=\bmatrix{0 & 1\\ 1&0},$ $\sigma_Z=\bmatrix{1&0\\ 0&-1},$ $\sigma_Y=\bmatrix{0 & -i \\ i & 0}$ are Pauli matrices. In PQCs, rotation gates are applied on individual qubits described by a unitary matrix \beano U (\theta_1, \theta_2,\hdots,\theta_n) &=& U_1(\theta_1)\otimes U_2(\theta_2)\otimes \cdots\otimes U_{n-1}(\theta_{n-1})\otimes U_n(\theta_n) \\ &=& (U_1(\theta_1)\otimes I\otimes\cdots \otimes I)(I\otimes U_2(\theta_2)\otimes \cdots \otimes I) \cdots (I\otimes I \otimes \cdots \otimes I \otimes U_n(\theta_n)) \\ &=& S_1(\theta_1)S_2(\theta_2)\cdots S_n(\theta_n)\eeano as illustrated in Figure \ref{fig:S.gate}. Also, recall that  \cite{horn1994topics} that 
\begin{equation}\label{eqn:expab}
    e^{A}\otimes e^B = e^{A\otimes I_m + I_n\otimes B}
\end{equation} for any matrices $A\in\C^{n\times n}, B\in\C^{m\times m}$ and $I_k$ denotes the identity matrix of order $k.$ In the following we derive generators of $U(\theta_1,\hdots,\theta_n).$    

%\begin{figure}[!ht]
%entering
%\begin{quantikz}
%& \gate{U_1} & \qw \\
%& \gate{U_2} & \qw \\
%& \vdots  & \\
%& \gate{U_{n-1}} &  \qw \\
%& \gate{U_{n}} &  \qw
%\end{quantikz} = \begin{quantikz}
%& \gate{U_1} \gategroup[wires=5,steps
%=1,style={inner sep=1pt}]{$S_1$} & \gate{I} \gategroup[wires=5,steps
%=1,style={inner sep=1pt}]{$S_2$}& \qw & \hdots & \gate{I}\gategroup[wires=5,steps
%=1,style={inner sep=1pt}]{$S_{n-1}$} & \gate{I}\gategroup[wires=5,steps
%=1,style={inner sep=1pt}]{$S_n$} & \qw \\
%& \gate{I} & \gate{U_2}& \qw & \hdots & \gate{I} & \gate{I} & \qw \\
%& \vdots  & \vdots & & \hdots & \vdots & \vdots & \\
%& \gate{I} & \gate{I} & \qw & \hdots & \gate{U_{n-1}}& \gate{I} & \qw \\
%& \gate{I} &  \gate{I} & \qw & \hdots & \gate{I} & \gate{U_n} & \qw
%\end{quantikz}
% \label{fig:S.gate}
%\end{figure} 

\begin{theorem}\label{thm:S.gate}
The unitary matrix represented by the quantum circuit Figure \ref{fig:S.gate} with $U_j\in \{R_X, R_Y, R_Z\},$ $1\leq j\leq n$
is given by $$U(\theta_1,\hdots,\theta_n)=\prod_{j=1}^n e^{-\iota \frac{1}{2}\theta_jH_j},$$ where $H_j=I_{2^{j-1}}\otimes \sigma_{s_j} \otimes I_{2^{n-j}}$ and  $U_j(\theta_j)=e^{-\iota \theta_j\frac{1}{2}\sigma_{s_j}},$ $s_j\in\{X,Y,Z\}.$ 

In particular, if  $U_j(\theta_j)=e^{-\iota \theta_j\frac{1}{2}\sigma_{s}},$ $s\in\{X,Y,Z\}$ for all $j$ then $$U(\theta_1,\hdots,\theta_n)= e^{-\iota \frac{1}{2} \sum_{j=1}^n \theta_jH_j}.$$
\end{theorem}
\pf Obviously, \beano S_j(\theta_j) &=& \underbrace{I\otimes \cdots \otimes I}_{(j-1)-times} \otimes U_j(\theta_j) \otimes \underbrace{I\otimes \cdots \otimes I}_{(n-j)-times} \\ &=& I_{2^{j-1}} \otimes e^{-\iota \theta_j\frac{1}{2}\sigma_{s_j}} \otimes I_{2^{n-j}} \\ &=& \frac{1}{e}e^{I_{2^{j-1}}} \otimes e^{-\iota \theta_j\frac{1}{2}\sigma_{s_j}} \otimes \frac{1}{e}e^{I_{2^{n-j}}} \\ &=& \frac{1}{e^2}e^{2I_{2^n} -\left(\iota \theta_j\frac{1}{2} I_{2^{j-1}}\otimes \sigma_{s_j} \otimes I_{2^{n-j}}\right)} \\ &=& e^{-\iota \theta_j\frac{1}{2} I_{2^{j-1}}\otimes \sigma_{s_j}\otimes I_{2^{n-j}}}. \eeano Now the desired result follows from the fact that $U(\theta_1,\hdots,\theta_n)=S_1(\theta_1)\cdots S_n(\theta_n).$  $\hfill{\square}$

Note that even if the Hamiltonians and the unitary matrices corresponding to $S_j(\theta_j),$ $1\leq j\leq n$ are easy to express and derive for small values of $n,$ the time complexity for finding the explicit matrix representation of $S_j(\theta_j)$ or $U(\theta_1,\hdots,\theta_n)$ is $O(2^{2n})$, using standard matrix multiplication. Now below we use the $\QT_n$ and show that this complexity be drastically reduced. We now consider a general one-qubit gate $U_j(\boldsymbol{\theta})=\bmatrix{e^{\iota(\theta_1+\theta_2)\cos\theta_3} & e^{\iota (\theta_1-\theta_2)}\sin\theta_3 \\ - e^{\iota (\theta_1-\theta_2)}\sin\theta_3 & e^{\iota (\theta_1+\theta_2)}\cos\theta_3 },$ where $\boldsymbol{\theta}=(\theta_1,\theta_2,\theta_3)\in\R^3.$ 
However, we denote $U_j=\bmatrix{u_{11} & u_{12} \\ u_{21} & u_{22}}$ for brevity in order to simplify the expression derived below.

Consider \begin{equation}\label{eqn:sj}S_j =I_2\otimes\hdots\otimes I_2\otimes \underbrace{U_j}_{j\mbox{-th position}}\otimes I_2\hdots \otimes I_2.\end{equation}
Then for any basis element $\ket{k_1\hdots  k_j\hdots k_n}$ of $\C^{2^n},$  \begin{equation}\label{cu1_0} S_j \ket{k_1\hdots k_j\hdots k_n}=\left\{
  \begin{array}{ll}
u_{11}\ket{k_1\hdots  k_{j-1} 0k_{j+1}\hdots k_n} + u_{21}\ket{k_1\hdots  k_{j-1} 1k_{j+1}\hdots k_n}, & \\   \hfill{\hbox{\mbox{if}\,\, $\ket{k_j}=\ket{0}$}} \\
u_{12}\ket{k_1\hdots k_{j-1} 0k_{j+1}\hdots k_n} + u_{22}\ket{k_1\hdots  k_{j-1} 1k_{j+1}\hdots k_n}, & \\  \hfill{\hbox{\mbox{if}\,\, $\ket{k_j}=\ket{1}.$}}
  \end{array}
\right.\end{equation}

Thus if $\ket{k_1\hdots k_i\hdots k_j\hdots k_n}$ is the $l$-th basis element in the standard ordered  basis of $\C^{2^n}$ then the corresponding output under $S_j$ can be written as a linear combination of $l$-th and $(l+2^{n-j})$-th basis elements if $\ket{k_j}=\ket{0},$ whereas the output is a linear sum of $l$-th and $(l-2^{n-j})$-th basis states if $\ket{k_j}=\ket{1}$ (note that if $k_j=1$ then $l> 2^{n-j}$), $1\leq l\leq 2^n.$ Then we have the following theorem. 

\begin{theorem}\label{thm:sgate}
The matrix representation of $S_j,$ defined in equation (\ref{eqn:sj}) is \begin{equation}\label{eqn:s_j}S_j=\diag\{\underbrace{\widehat{U}, \hdots, \widehat{U}}_{2^{j-1}\mbox{-times}}\},\end{equation} where $\widehat{U}=\bmatrix{\widehat{u}_{pq}}\in\C^{2^{n-j+1}\times 2^{n-j+1}}$ is given by 
\begin{equation}\label{eqn:upq_0}
\widehat{u}_{pq}=\left\{
  \begin{array}{ll}
u_{11} & \hbox{$1\leq p=q \leq 2^{n-j}$} \\
u_{12} & \hbox{$p=r, q=r+2^{n-j}, r\in\{1,\hdots,2^{n-j}\}$}\\
u_{21} & \hbox{$p=r+2^{n-j}, q=r, r\in \{1,\hdots,2^{n-j}\}$}\\
u_{22} & \hbox{$1+2^{n-j}\leq p= q\leq 2^{n-j+1}$}\\
0 & \hbox{otherwise.}
\end{array}
\right.\end{equation}
Finally, the Hamiltonian $H_j$ such that $S_j=e^{-\iota H_j}$ is given by equation \ref{eqn:hsg}.
\end{theorem}

\pf Note that $S_j$ acts on the basis elements as described by equation (\ref{eqn:upq_0}) and an output state depends on the $j$-th qubit of the corresponding input state. From the construction of $\QT_n,$ it follows that there are $2^{n-j}$ terminal nodes stem from each $j$-th order node. The last qubit of the state corresponding to a $j$-th order state determine the $j$-th qubit state of an $n$-qubit basis state represented by a terminal node of $\QT_n.$ Since there are alternative $\ket{0}$ and $\ket{1}$ last-qubit state in the states corresponding to the $j$-th order nodes of $\QT_n,$ and there are $2^{n-j}$ $n$-qubit basis states originated from each of these $j$-th order nodes, $S_j$ acts alternatively on $j$-th qubit as $\ket{0}$ and $\ket{1}$ of the $n$-qubit basis states in groups of $2^{n-j}$, starting from the group of $2^{n-j}$ basis elements whose $j$-th qubit is $\ket{0}.$ Thus the proof of equation (\ref{eqn:s_j}) follows. The equation (\ref{eqn:upq_0}) follows by employing equation (\ref{cu1_0}). Observe that the $S_j$ has a similar form as that of equation (\ref{eqn:uhatc1}). 

Now following the proof of Theorem \ref{thm:ilj} for finding the eigenpairs of the matrix given in equation (\ref{eqn:uhatc1}), the eigenpairs of $S_j$ can be obtained from eigenpairs of $U_j$ as follows. Let $(\lambda_s, \ket{u_s}= \bmatrix{\widehat{u}_{s1} & \widehat{u}_{s2}}^T),$ $s=1,2$ be the eigenpairs of $U.$ Then for any $r\in\{1,\hdots,2^{n-j}\},$ the vector $\ket{v_r}_s=[v_{s,r}]_{l=1}^{2^{n-j+1}}\in \C^{2^{n-j+1}}$ given by 
\begin{equation}\label{eqn:upq1}
[v_{s,r}]_l=\left\{
  \begin{array}{ll}
\widehat{u}_{s1} & \hbox{if $l=r$} \\
\widehat{u}_{s2} & \hbox{if $l=r+2^{n-j}$}\\
0 & \hbox{otherwise.}
\end{array}
\right.\end{equation} is an eigenvector of $\widehat{U}$ corresponding to the eigenvalue $\lam_s, s=1,2.$ Thus $\left(\lam_s,\ket{v_r}_s\right), s=1,2,$  $r\in\{1,\hdots,2^{n-j}\}$ provide a complete list of eigenpairs of $\widehat{U}.$ 

Then if $\{\ket{e_m}: 1\leq m\leq 2^{j-1}\}$ is the canonical basis of $\C^{2^{j-1}}$ then $(\lam_s,\ket{e_m}\otimes \ket{v_r}_s)$ form a complete list of eigenpairs of $S_j,$ where $s=1,2,$  $r\in\{1,\hdots,2^{n-j}\}$ and $m\in\{1,\hdots, 2^{j-1}\}.$ Hence the eigendecomposition of $S_j$ is given by $$S_j=\sum_{s,r,m} \lam_s (\ket{e_m}\otimes \ket{v_r}_s) (\bra{e_m}\otimes \bra{v_r}_s).$$ 

Finally, by Proposition \ref{prop:eigexp}, the Hamiltonian $H_j$ such that $S_j=e^{-\iota H_j}$ is given by \begin{equation}\label{eqn:hsg} H_j=\sum_{s,r,m} z_s (\ket{e_m}\otimes \ket{v_r}_s) (\bra{e_m}\otimes \bra{v_r}_s),\end{equation} where $\lam_s=e^{-\iota z_s},$ $s=1,2.$ This completes the proof. \hfill{$\square$}

Thus for a string of one-qubit gates $U(\boldsymbol{\theta})=\prod_{j=1}^n S_j(\boldsymbol{\theta}_j)$, the local Hamiltonian decomposition of $U(\boldsymbol{\theta})$ is given by $$U(\boldsymbol{\theta})=\prod_{j=1}^n e^{-\iota H_j},$$ where $H_j$ is given by Theorem \ref{thm:sgate}.

\subsubsection{Numerical Simulations}

Now we validate Theorem \ref{thm:sgate} for certain  standard string of one-qubit gates in a PQC  for calculation of Hamiltonian with direct computation of exponential of a matrix using \textit{Python} $3$'s \textit{mexp} function. We assume a $4$-qubit circuit the string of roation gates $R_U,$ $U\in \{R_X(\theta), R_Y(\theta), R_Z(\theta)\}$ applied on each qubit. Then using the decomposition in Figure \ref{fig:S.gate} setting $U_j=U,$ we calculate the error $\|U(\boldsymbol{\theta}) - e^{\iota H_1(\theta)} e^{-\iota H_2} e^{\iota H_3(\theta)} e^{\iota H_4(\theta)}\|_F$ for $100$ values of $\theta\in [-\pi, \pi].$ The $H_j$ is found employing Theorem \ref{thm:sgate} and $U(\boldsymbol{\theta})$ is obtained by finding the tensor product of the rotation gates in \textit{Python} $3.$ The errors are in the order of $10^{-15}.$

\begin{figure}[h]
    \centering
    \includegraphics[width=0.5\linewidth]{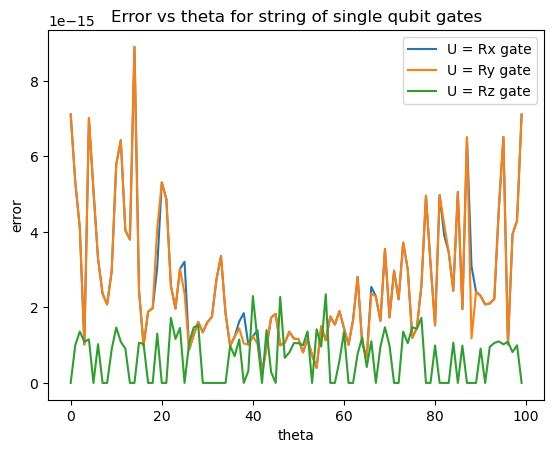}
    \caption{The error $\|U(\boldsymbol{\theta}) -e^{\iota H_1(\theta)} e^{-\iota H_2} e^{\iota H_3(\theta)} e^{\iota H_4(\theta)} \|_F$ vs $\theta$ for $100$ values of $\theta\in [-\pi, \pi]$ }
    \label{fig:enter-label}
\end{figure}

\subsection{Local Hamiltonian decomposition}

Given a one-layer $n$-qubit PQC defined by single-qubit and two-qubit control parametriz gates, which represents a unitary matrix $U(\boldsymbol{\theta}),$ following the description as in Figure \ref{fig:PQC} we have \begin{equation}\label{eqn:uqg}U(\boldsymbol{\theta})=\prod_{k=1}^K U_k(\boldsymbol{\theta}_k),\end{equation} where $U_k(\boldsymbol{\theta}_k)$ represents a string of single-qubit parametrized gates or a two-qubit parametrized control gate. Here $\boldsymbol{\theta}=\{\boldsymbol{\theta_{k}} : 1\leq k\leq K\}$ and $\boldsymbol{\theta}_k$ denotes the set of parameters involved in the string of single-qubit gates or two-qubit control gates.

Then following Theorem \ref{thm:ilj} and Theorem \ref{thm:igj}, utilizing the Hamiltonian representation, we have \begin{equation}\label{eqn:tqg}U_k(\boldsymbol{\theta}_k)=e^{-\iota  \sum_{p=1}^P \lam_{pk}(\boldsymbol{\theta}_k)H_{pk} (\boldsymbol{\theta}_k)}\end{equation} corresponding to a two-qubit control gate, where $P$ is the number of non-unit eigenvalues of $U_k(\boldsymbol{\theta}_k).$

Following Theorem \ref{thm:sgate}, when $U_k(\boldsymbol{\theta}_k)$ represents a string of single-qubit gates then \begin{equation}\label{eqn:sqg}U_k(\boldsymbol{\theta}_k)=\prod_{j=1}^n S_j(\boldsymbol{\theta}_{jk})=\prod_{j=1}^n e^{-\iota \sum_{q=1}^Q \lambda_{qk}(\boldsymbol{\theta}_{qk})H_{qk}(\boldsymbol{\theta}_k)},\end{equation} where $\boldsymbol{\theta}_{jk}$ denotes the set of parameters in the $k$-th single-qubit gate of the string of single-qubit gates which define $U_k(\boldsymbol{\theta}_k),$ and $Q$ is the number of non-unit eigenvalues of $U_k(\boldsymbol{\theta}_k).$   

Thus employing equations (\ref{eqn:tqg}) and (\ref{eqn:sqg}) into equation (\ref{eqn:uqg}) gives the local Hamiltonian decomposition of the PQC.  

\section{Probability amplitudes of the output state of a quantum circuit}\label{sec3}

Finally, we use the decomposition of a single layer PQC as described in Figure \ref{fig:PQC} and derive analytical formula for computation of probability amplitudes of output of an input state under the given PQC. Obviously, the unitary matrix corresponding to the PQC can be writtn as a product of unitaries corresponding to string of one-qubit unitaries and two-qubit control gates. Employing the sparsity representation of these local unitaries, we develop fast algorithms for the determination of the probability amplitudes corresponding to canonical basis for the output state corresponding to an input state for a PQC with one-qubit and two-qubit control gates. 

In order to derive the output state for an input state under a two-qubit control gate $C_{ij}$ with $j$-th qubit as target and $i$-th qubit as control in a $n$-qubit system, let us consider $i<j$ and the unitary matrix corresponding to $C_{ij}$ given by equations (\ref{cum}) and (\ref{eqn:uhatc1}). 
Let $\ket{\psi}=\sum_{k=1}^{2^n} a_k \, \ket{k}$ be an input state, where $\{\ket{k}: \, 1\leq k\leq 2^n\}$ denotes the canonical ordered basis of $\C^{2^n}.$ Then 

\begin{eqnarray}
&& C_{ij}\ket{\psi} =  \sum_{l=0}^{{2^{i-1}-1}} \sum_{k=2l\times 2^{n-i}+1}^{(2l+1)\times 2^{n-i}} a_k \, \ket{k}  + \nonumber \\
&& \sum_{l=0}^{2^{i-1}-1} \left[ 
 (u_{11}a_{(2l+1)2^{n-i}+1}+u_{12}a_{(2l+1)2^{n-i}+2^{n-j}+1})\, \ket{(2l+1)2^{n-i}+1} + \hdots+ \right. \nonumber \\
&& (u_{11}a_{(2l+1)2^{n-i}+2^{n-j}}+u_{12}a_{(2l+1)2^{n-i}+2^{n-j+1}})\, \ket{(2l+1)2^{n-i}+2^{n-j}} + \nonumber \\
&& (u_{21}a_{(2l+1)2^{n-i}+1}+u_{22}a_{(2l+1)2^{n-i}+2^{n-j}+1})\, \ket{(2l+1)2^{n-i}+2^{n-j}+1} +\hdots+ \nonumber \\
&& (u_{21}a_{(2l+1)2^{n-i}+2^{n-j}}+u_{22}a_{(2l+1)2^{n-i}+2^{n-j+1}})\, \ket{(2l+1)2^{n-i}+2^{n-j+1}} + \nonumber \\
&& (u_{11}a_{2^{n-j+1}+(2l+1)2^{n-i}+1} + u_{12}a_{(2^{j-i-1}-1)2^{n-j+1}+(2l+1)2^{n-i}+2^{n-j}+1})\, \ket{2^{n-j+1}+(2l+1)2^{n-i}+1} + \hdots + \nonumber \\
&& (u_{11}a_{2^{n-j+1}+(2l+1)2^{n-i}+2^{n-j}} + u_{12}a_{2^{n-j+1}+(2l+1)2^{n-i}+2^{n-j+1}})\, \ket{2^{n-j+1}+(2l+1)2^{n-i}+2^{n-j}} + \nonumber \\
&& (u_{21}a_{2^{n-j+1}+(2l+1)2^{n-i}+1} + u_{22}a_{2^{n-j+1}+(2l+1)2^{n-i}+2^{n-j}+1})\, \ket{2^{n-j+1}+(2l+1)2^{n-i}+2^{n-j}+1} + \hdots + \nonumber \\
&& (u_{21}a_{2^{n-j+1}+(2l+1)2^{n-i}+2^{n-j}} + u_{22}a_{2^{n-j+1}+(2l+1)2^{n-i}+2^{n-j+1}})\, \ket{2^{n-j+1}+(2l+1)2^{n-i}+2^{n-j+1}} + \nonumber \\
&& \vdots \nonumber \\
&& (u_{11}a_{(2^{j-i-1}-1)2^{n-j+1}+(2l+1)2^{n-i}+1} + u_{12}a_{(2^{j-i-1}-1)2^{n-j+1}+(2l+1)2^{n-i}+2^{n-j}+1})\, \nonumber \\ 
&& \hfill{\ket{(2^{j-i-1}-1)2^{n-j+1}+(2l+1)2^{n-i}+1} + \hdots +} \nonumber \\
&& (u_{11}a_{(2^{j-i-1}-1)2^{n-j+1}+(2l+1)2^{n-i}+2^{n-j}} + u_{12}a_{(2^{j-i-1}-1)2^{n-j+1}+(2l+1)2^{n-i}+2^{n-j+1}})\, \nonumber \\
&& \hfill{\ket{(2^{j-i-1}-1)2^{n-j+1}+(2l+1)2^{n-i}+2^{n-j}} +} \nonumber \\
&& (u_{21}a_{(2^{j-i-1}-1)2^{n-j+1}+(2l+1)2^{n-i}+1} + u_{22}a_{(2^{j-i-1}-1)2^{n-j+1}+(2l+1)2^{n-i}+2^{n-j}+1})\, \nonumber \\
&& \hfill{\ket{(2^{j-i-1}-1)2^{n-j+1}+(2l+1)2^{n-i}+2^{n-j}+1} + \hdots +} \nonumber \\
&& (u_{21}a_{(2^{j-i-1}-1)2^{n-j+1}+(2l+1)2^{n-i}+2^{n-j}} + u_{22}a_{(2^{j-i-1}-1)2^{n-j+1}+(2l+1)2^{n-i}+2^{n-j+1}}) \nonumber \\ 
&& \left. \hfill{\ket{(2^{j-i-1}-1)2^{n-j+1}+(2l+1)2^{n-i}+2^{n-j+1}} } \right]. \label{eqn:cuilj}
\end{eqnarray}

Note that $C_{ij}$ (given by $C_U$ in equation (\ref{cum})) acts as an identity operator on the alternative $2^{n-i}$ standard basis elements, the coefficients corresponding to these basis states in the algebraic vector representation of $\ket{\psi}$ remain invariant under the gate $U_{ij}.$ Thus the first line of the expression of $U_{ij}\ket{\psi}$ follows. Now for the basis states on which $C_U$ acts nontrivially through the expression of $\widehat{U},$ which is further defined by the blocks $\widehat{U}^{(1)}\in\C^{2^{n-j+1}\times 2^{n-j+1}}$ acts symmetrically on $2^{n-j+1}$ basis states following equation (\ref{cu1}) and equation (\ref{eqn:upq}). For example, the effect of the first block $\widehat{U}^{(1)}$ in the expression of the first block $\widehat{U}$ of $C_U$ given by equation (\ref{eqn:uhatc1}) on the input state $\ket{\psi}$ as described above is given by 
\begin{eqnarray*}
 &&   (u_{11}a_{2^{n-i}+1}+u_{12}a_{2^{n-i}+2^{n-j}})\, \ket{2^{n-i}+1} + \hdots+(u_{11}a_{2^{n-i}+2^{n-j}}+u_{12}a_{2^{n-i}+2^{n-j+1}})\, \ket{2^{n-i}+2^{n-j}} + \\
&& (u_{21}a_{2^{n-i}+1}+u_{22}a_{2^{n-i}+2^{n-j}})\, \ket{2^{n-i}+2^{n-j}+1} +\hdots+ (u_{21}a_{2^{n-i}+2^{n-j}}+u_{22}a_{2^{n-i}+2^{n-j+1}})\, \ket{2^{n-i}+2^{n-j+1}} 
\end{eqnarray*}
setting $l=0.$ Thus the remaining equations in the output state $C_U\ket{\psi}$ represent the action of a block matrix $\widehat{U},$ where $l$ represents the position of the block in the matrix representation of $C_U.$

\begin{example}
    \begin{itemize}
        \item[(a)] Consider the example given in Example \ref{exp1} (a). If $\ket{\psi}=\sum_{k=1}^{32} a_k\ket{k}$, where $\{\ket{k}: 1\leq k\leq 32\}$ is the standard basis of $\C^{2^5}$ then 
\begin{eqnarray*}
&& U_{ij}\, \ket{\psi} = \sum_{k=1}^{8} a_k \ket{k} + \\ 
&& (u_{11}a_9+u_{12}a_{13}) \ket{9} +(u_{11}a_{10}+u_{12}a_{14}) \ket{10} + (u_{11}a_{11}+u_{12}a_{15}) \ket{11} +  (u_{11}a_{12}+u_{12}a_{16}) \ket{12} + \\
&& (u_{21}a_{9}+u_{22}a_{13}) \ket{13} + (u_{21}a_{10}+u_{22}a_{14}) \ket{14} +(u_{21}a_{11}+u_{22}a_{15}) \ket{15}  + (u_{21}a_{12}+u_{22}a_{16}) \ket{16} + \\
&& \sum_{k=17}^{24} a_k\ket{k} + \\
&& (u_{11}a_{25}+u_{12}a_{29}) \ket{25} +(u_{11}a_{26}+u_{12}a_{30}) \ket{26} + (u_{11}a_{27}+u_{12}a_{31}) \ket{27} +  (u_{11}a_{28}+u_{12}a_{32}) \ket{28} + \\
&& (u_{21}a_{25}+u_{22}a_{29}) \ket{29} + (u_{21}a_{26}+u_{22}a_{30}) \ket{30} +(u_{21}a_{27}+u_{22}a_{31}) \ket{31}  + (u_{21}a_{28}+u_{22}a_{32}) \ket{32}.
\end{eqnarray*}

\item[(b)] Consider the input state $\ket{\psi}=\sum_{k=1}^{32} a_k\, \ket{k}$ given in Example \ref{exp1} (b). Them the corresponding output state is given by
\begin{eqnarray*}
&& U_{ij}\, \ket{\psi} = \sum_{k=1}^{8} a_k \ket{k} + \\ 
&& (u_{11}a_9+u_{12}a_{11}) \ket{9} +(u_{11}a_{10}+u_{12}a_{12}) \ket{10} + (u_{21}a_{9}+u_{22}a_{11}) \ket{11} +  (u_{21}a_{10}+u_{22}a_{12}) \ket{12} + \\
&& (u_{11}a_{13}+u_{12}a_{15}) \ket{13} + (u_{11}a_{14}+u_{12}a_{16}) \ket{14} +(u_{21}a_{13}+u_{22}a_{15}) \ket{15}  + (u_{21}a_{14}+u_{22}a_{16}) \ket{16} + \\
&& \sum_{k=17}^{24} a_k\ket{k} + \\
&& (u_{11}a_{25}+u_{12}a_{27}) \ket{25} +(u_{11}a_{26}+u_{12}a_{28}) \ket{26} + (u_{21}a_{25}+u_{22}a_{27}) \ket{27} +  (u_{21}a_{26}+u_{22}a_{28}) \ket{28} + \\
&& (u_{11}a_{29}+u_{12}a_{31}) \ket{29} + (u_{11}a_{30}+u_{12}a_{32}) \ket{30} +(u_{21}a_{29}+u_{22}a_{31}) \ket{31}  + (u_{21}a_{29}+u_{22}a_{32}) \ket{32}.
\end{eqnarray*}

    \end{itemize}
\end{example}

Similarly, we can obtain direct formulae for probability amplitudes for $C_{ij}\ket{\psi}$ when $i>j$, and for output state of an input under a string of one-qubit gates. However, it is not immediate what gain is achieved from these expressions. Thus we develop algorithms for the generation of the output states corresponding to an input state when it undergoes these gates in a quantum circuit setup. Then we analyze the complexity of these algorithms.   

%Now incorporating the approach proposed in Figure .. we can derive the probability amplitudes of the output of any $n$-qubit input state for a quantum circuit comprising one-qubit and two-qubit control gates can be derived as follows. 

First, we provide Algorithm(\ref{alg:cpg}) in order to determine output states of an input state for a two-qubit control gate. Next, in Algorithm \ref{alg:sqg} we determine output state for a string of one-qubit gates. Note that these algorithms exploit the sparsity pattern of the $2$-sparse unitary matrices of order $2^n$ corresponding to single-qubit and two-qubit control gates derived in Theorem \ref{thm:ilj}, Theorem \ref{thm:igj} and Theorem \ref{thm:sgate}.

\begin{algorithm}
\caption{Algorithm for applying a control gate on a quantum state}
\label{alg:cpg}
\begin{algorithmic}[1]
\Require The fixed variables n,i,j and $1 \le i,j \le n$ , $i \ne j$. An input quantum state $inp \in \C^{2^{n}}$. A unitary matrix $U=\bmatrix{u_{11} & u_{12} \\ u_{21} & u_{22}}.$
\Ensure Quantum State after the application of control gate $out \in \C^{2^{n}}$
\Procedure{ControlGate}{$\text{inp}, n, i, j, U$}
    \State $out \gets inp$
    \If{$i < j$}
        \State $state \gets 1$
        \For{$x \gets 1$ to $2^i$}
            \If{$x$ is odd}
                \State $state \gets state + 2^{(n-i)}$
            \EndIf
            \If{$x$ is even}
                \For{$a \gets 1$ to $2^{(j-i-1)}$}
                    \For{$b \gets 1$ to $2^{(n-j)}$}
                        \State $\text{out}_{state} \gets u_{11} \cdot \text{inp}_{state} + u_{12} \cdot \text{inp}_{state+2^{(n-j)}}$
                        \State $state \gets state + 1$
                    \EndFor
                    \For{$b \gets 1$ to $2^{(n-j)}$}
                        \State $\text{out}_{state} \gets u_{21} \cdot \text{inp}_{state-2^{(n-j)}} + u_{22} \cdot \text{inp}_{state}$
                        \State $state \gets state + 1$
                    \EndFor
                \EndFor
            \EndIf
        \EndFor
    \Else
        \For{$r \gets 1$ to $2^{(j-1)}$}  % r <- 1
            \For{$l \gets (2r-2) \cdot 2^{(i-j-1)}$ to $(2r-1) \cdot 2^{(i-j-1)}$}
                \For{$k \gets (2l+1) \cdot 2^{(n-i)}+1$ to $(2l+2) \cdot 2^{(n-i)}$}
                    \State $\text{out}_{k} \gets u_{11} \cdot \text{inp}_{k} + u_{12} \cdot \text{inp}_{k+2^{(n-j)}}$
                \EndFor
            \EndFor
        \EndFor
        \For{$r \gets 1$ to $2^{(j-1)}$}
            \For{$l \gets (2r-1) \cdot 2^{(i-j-1)}$ to $(2r) \cdot 2^{(i-j-1)}$}
                \For{$k \gets (2l+1) \cdot 2^{(n-i)}+1$ to $(2l+2) \cdot 2^{(n-i)}$}
                    \State $\text{out}_{k} \gets u_{21} \cdot \text{inp}_{k-2^{(n-j)}} + u_{22} \cdot \text{inp}_{k}$
                \EndFor
            \EndFor
        \EndFor
    \EndIf
    \State \textbf{return} $out$
\EndProcedure
\end{algorithmic}
\end{algorithm}

\begin{algorithm}
    \caption{Algorithm for applying a single qubit gate on a quantum state}
    \label{alg:sqg}
    \begin{algorithmic}
        \Require The fixed variables $n,j$. An input quantum state $inp \in \C^{2^{n}}$. A unitary matrix $U=\bmatrix{u_{11} & u_{12} \\ u_{21} & u_{22}}.$
        \Ensure Quantum State after the application of single qubit gate $out \in \C^{2^{n}}$
        \Procedure{SingleQubitGate}{$\text{inp}, n, j, U$}
        \State $out \gets inp$
        \State $state \gets 1$
        \For{$x \gets 1$ to $2^{j}$}
            \If{$x$ is odd}
                \For{$a \gets 1$ to $2^{(n-j)}$}
                    \State $out_{state} \gets u_{11} \cdot \text{inp}_{state} + u_{12} \cdot \text{inp}_{state + 2^{(n-j)}}$
                    \State $state \gets state + 1$
                \EndFor
            \EndIf
            \If{$x$ is even}
                \For{$a \gets 1$ to $2^{(n-j)}$}
                    \State $out_{state} \gets u_{21} \cdot \text{inp}_{state - 2^{(n-j)}} + u_{22} \cdot \text{inp}_{state}$
                    \State $state \gets state + 1$
                \EndFor
            \EndIf
        \EndFor
        \State \textbf{return} $out$
        \EndProcedure
    \end{algorithmic}
\end{algorithm}

\subsection{Complexity of the algorithms}

%\textbf{For Single Qubit gate:} \newline
% To apply a single qubit gate U on qubit j in an n qubit circuit with initial state $\ket{\psi}$ we first need to calculate 
% $$S_j = \underbrace{I\otimes \cdots \otimes I}_{(j-1)-times} \otimes U \otimes \underbrace{I\otimes \cdots \otimes I}_{(n-j)-times}$$
% $S_j$ is $\C^{2^{n} \times 2^{n}}$ and $\ket{\psi}$ is $\C^{2^{n}}$ calculating $\ket{\psi}' = S_{j}\psi$ takes $O((2^{n})^{2})$ operations and the construction of $S_j$ takes $O(2^{n})$ \cite{Vidal_Romero_2023} hence the total operations will be $O(2^{2n})$+$O(2^{n})$ and time complexity of the operation will be $O(2^{2n})$\newline

To apply a single-qubit gate using Algorithm(\ref{alg:sqg}) we need to iterate over all $2^{n}$ values in the input quantum state and to update every state we need $O(1)$ operations. Thus  the time complexity of the Algorithm \ref{alg:sqg} is $O(2^{n}).$ Next, to apply a control gate using Algorithm (\ref{alg:cpg}) we need to iterate over all $2^{n}$ values in the input quantum state and it either remains the same or changes in $O(1)$ number of entries. Thus the time complexity of this  algorithm is $O(K\, 2^{n}),$ where $K$ is the total number of single-qubit and two-qubit control gates.

\begin{figure}[h]
    \centering
         \subfloat [\centering ]{{\includegraphics[width=0.45\textwidth]{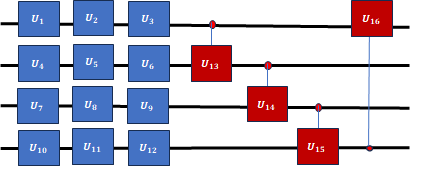} }}%
    \qquad
    \subfloat [\centering ]{{\includegraphics[width=0.45\textwidth]{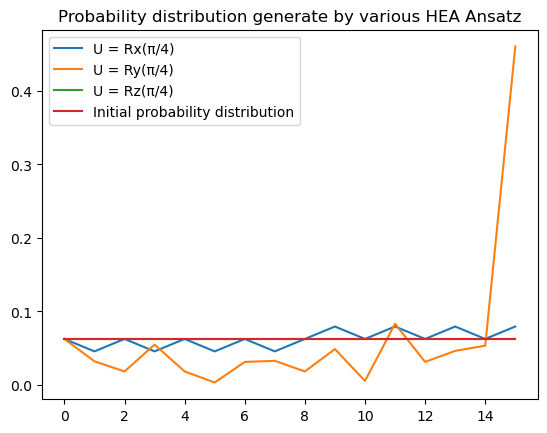} }}%
    \caption{(a)$4$-qubit Hardware Efficient Ansatz (HEA), $U_j=R_X(\theta_j),$ $j=1,\hdots,16$ (b) Probability distribution generated by HEA Ansatz using different controlled rotation gates applied using Algorithm(\ref{alg:cpg}) setting $U_j(\theta_j)=U\in\{R_X(\pi/4), R_Y(\pi/4), R_Z(\pi/4)\}.$ }
    \label{fig:HEA}
\end{figure}

\subsection{Numerical simulations}

We consider the $4$-qubit Hardware Efficient Ansatz (HEA) (Figure \ref{fig:HEA} (a) ) to generate probability distributions corresponding to the output state for the input state which corresponds to uniform distribution. 

%\begin{figure}
 %   \centering
  %  \includegraphics[width=0.45\linewidth]{fig_heag.png}
   % \caption{$4$-qubit Hardware Efficient Ansatz (HEA), $U_j=R_X(\theta_j),$ $j=1,\hdots,16$}
    %\label{fig:HEA}
%\end{figure}

As discussed in Figure \ref{fig:PQC}, the unitary matrix $U(\boldsymbol{\theta}),$ $\boldsymbol{\theta}=\{\theta_j : 1\leq j\leq 16\}$ corresponding to the $4$-qubit HEA can be written as a product of $6$ local unitaries, the first three of which correspond to strings of one-qubit gates and the last three are two-qubit control unitaries. Then setting $U_j$ to be one of the fixed rotation gates $R_X(\theta_j),$ $R_Y(\theta_j),$ $R_Z(\theta_j)$ we plot the probability distribution of the output states by setting $\theta_j=\pi/4$ for all $j$, employing Algorithms \ref{alg:cpg} and \ref{alg:sqg} in Figure \ref{fig:HEA} (b). \\

%In the Figure(\ref{fig:outdist}) input is a state with equal probability of each quantum state i.e. uniform distribution of states and then controlled rotational gates are applied with target gate $U$, following controlled gate are used in order $1-2, 2-3, 3-2, 4-1$ where for $i-j$, i is the control qubit and j is the target qubit. \newline If we take $U = X$ as the target gate with input state as uniform and if the control is on $i^{th}$ qubit and target is $j^{th}$ qubit and the probability of state $\ket{\hdots1_{i}\hdots0_{j}\hdots}$ and $\ket{\hdots1_{i}\hdots1_{j}\hdots}$ are equal and after the application of gate the states become $\ket{\hdots1_{i}\hdots1_{j}\hdots}$ and $\ket{\hdots1_{i}\hdots0_{j}\hdots}$ respectively hence the probability distribution remains the same. \newline If we take $U = R_{Z}$ where all states have equal probability and let the state be s, then by Algorithm(\ref{alg:cpg}) we can see that the state s can be transformed to $s' = se^{i\theta/2}$ or $s' = se^{-i\theta/2}$ from here we can see that $ss^{*}$ and $s's'^{*}$ are equal. hence the probability distribution will not change and it will remain equiprobable.

%\begin{figure}
  %  \centering
   % \includegraphics[width=0.5\linewidth]{outdist_HEA.png}
    %\caption{Probability distribution generated by HEA Ansatz using different controlled rotation gates applied using Algorithm(\ref{alg:cpg})}
    %\label{fig:outdist}
%\end{figure}

\noindent{\bf Conclusion.} In this paper we undertake a detailed study on finding unitary matrices of order $2^n$ for single-qubit and two-qubit control gates in any parametrized quantum circuit (PQC) of $n$-qubits. We associate the nodes of the complete binary one-rooted tree as the canonical ordered basis elements of any $k$-qubit Hilbert space by the $k$-th order nodes of the tree, $1\leq k\leq n$. This combinatorial representation enables us to identify and derive each entry of the desired unitary matrices and we show that these matrices are $2$-sparse. Then using a linear algebraic argument, we derive Hamiltonians corresponding to these gates, that we call local Hamiltonian for the entire PQC. Consequently, a local Hamiltonian decomposition of the PQC is obtained. Further we employ the sparsity pattern to develop a classical algorithm of complexity $O(2^n)$ to determine probability amplitudes of the output of a PQC corresponding to any input quantum state.     \\

\noindent{\bf Acknowledgement.} AJ thanks Fujitsu Research of India Pvt Ltd (FRIPL) for the research internship opportunity to work on this project.

\bibliographystyle{plain}
\bibliography{reff}

\end{document}